\documentclass[11pt]{article}
\usepackage[utf8]{inputenc}

\usepackage{amsmath,amssymb,amsthm}
\usepackage{multirow}
\usepackage{tikz}
\usepackage{float}
\usetikzlibrary{calc}
\usepackage[margin=1in]{geometry}
\usepackage{xcolor}
\definecolor{oblue}{RGB}{0, 50, 120}
\usepackage{color}
\usepackage{xspace}
\usepackage{algorithm,algpseudocode}
\usepackage{bm}
\usepackage[hyperfootnotes=false, colorlinks=true, linkcolor=black, urlcolor=black, citecolor={blue}]{hyperref}
\usepackage{soul}
\usepackage{framed}
\usepackage{caption}
\usepackage{subcaption}
\usepackage{tikz}
\usepackage{tikz-cd}
\usepackage{enumerate}
\usepackage{enumitem}
\setitemize{noitemsep,topsep=0pt,parsep=0pt,partopsep=0pt}
\usepackage{listings}
\usepackage{booktabs}

\colorlet{punct}{red!60!black}
\definecolor{background}{HTML}{EEEEEE}
\definecolor{delim}{RGB}{20,105,176}
\colorlet{numb}{magenta!60!black}

\lstdefinelanguage{json}{
    basicstyle=\normalfont\ttfamily,
    numbers=left,
    numberstyle=\scriptsize,
    stepnumber=1,
    numbersep=8pt,
    showstringspaces=false,
    breaklines=true,
    frame=lines,
    backgroundcolor=\color{background},
    literate=
     *{0}{{{\color{numb}0}}}{1}
      {1}{{{\color{numb}1}}}{1}
      {2}{{{\color{numb}2}}}{1}
      {3}{{{\color{numb}3}}}{1}
      {4}{{{\color{numb}4}}}{1}
      {5}{{{\color{numb}5}}}{1}
      {6}{{{\color{numb}6}}}{1}
      {7}{{{\color{numb}7}}}{1}
      {8}{{{\color{numb}8}}}{1}
      {9}{{{\color{numb}9}}}{1}
      {:}{{{\color{punct}{:}}}}{1}
      {,}{{{\color{punct}{,}}}}{1}
      {\{}{{{\color{delim}{\{}}}}{1}
      {\}}{{{\color{delim}{\}}}}}{1}
      {[}{{{\color{delim}{[}}}}{1}
      {]}{{{\color{delim}{]}}}}{1},
}

\usepackage{bbm}

\usepackage{setspace}
\newcommand{\inparen}[1]{\left( #1 \right) }
\newcommand{\inbrak}[1]{\left[ #1 \right] }

\newcommand{\inset}[1]{\left\{ #1 \right\} }

\newcounter{protocol}

\usepackage{natbib}
 \bibpunct[, ]{(}{)}{,}{a}{}{,}%
\newcommand{\BlurTimeMin}{04:49, October 19th, 2022\xspace}
\newcommand{\BlurTimeMax}{17:14, March 7th, 2024\xspace}
\newcommand{\BlurTradeCounts}{3503420\xspace}
\newcommand{\BlurTotalVol}{3507602.25\xspace}
\newcommand{\BlurCollectionCount}{7495\xspace}
\newcommand{\BlurSalesNinetynine}{3319.01\xspace}
\newcommand{\BlurMedpriceMax}{187.0\xspace}
\newcommand{\BlurMedpriceNinetynine}{7.37\xspace}
\newcommand{\LooksRareTimeMin}{11:04, December 29th, 2021\xspace}
\newcommand{\LooksRareTimeMax}{07:56, April 13th, 2023\xspace}
\newcommand{\LooksRareTradeCounts}{401636\xspace}
\newcommand{\LooksRareTotalVol}{10026162.88\xspace}
\newcommand{\LooksRareCollectionCount}{9474\xspace}
\newcommand{\LooksRareSalesNinetynine}{741.78\xspace}
\newcommand{\LooksRareMedpriceMax}{11307.24\xspace}
\newcommand{\LooksRareMedpriceNinetynine}{19.13\xspace}
\newcommand{\OpenSeaTimeMin}{19:57, February 18th, 2022\xspace}
\newcommand{\OpenSeaTimeMax}{13:46, March 20th, 2024\xspace}
\newcommand{\OpenSeaTradeCounts}{21514100\xspace}
\newcommand{\OpenSeaTotalVol}{5910280.56\xspace}
\newcommand{\OpenSeaCollectionCount}{42442\xspace}
\newcommand{\OpenSeaSalesNinetynine}{1759.13\xspace}
\newcommand{\OpenSeaMedpriceMax}{2000.0\xspace}
\newcommand{\OpenSeaMedpriceNinetynine}{2.74\xspace}
\newcommand{\BlurFiloneCount}{5239\xspace}
\newcommand{\BlurFiloneFrac}{0.15}
\newcommand{\BlurFiloneVol}{0.36}
\newcommand{\BlurFiltwoCount}{252204\xspace}
\newcommand{\BlurFiltwoFrac}{7.2}
\newcommand{\BlurFiltwoVol}{27.2}
\newcommand{\BlurFilthreeCount}{181240\xspace}
\newcommand{\BlurFilthreeFrac}{5.17}
\newcommand{\BlurFilthreeVol}{23.56}
\newcommand{\BlurFilfourCount}{1244360\xspace}
\newcommand{\BlurFilfourFrac}{35.52}
\newcommand{\BlurFilfourVol}{38.47}
\newcommand{\BlurFilallfourCount}{1431537\xspace}
\newcommand{\BlurFilallfourFrac}{40.86}
\newcommand{\BlurFilallfourVol}{58.16}
\newcommand{\BlurNWmedpriceMed}{0.0131\xspace}
\newcommand{\BlurWmedpriceMed}{0.0189\xspace}
\newcommand{\BlurWashtraderCount}{115174\xspace}
\newcommand{\BlurWashgroupCount}{2007\xspace}

\newcommand{\BlurNWprofitMedian}{0.014515\xspace}
\newcommand{\BlurWprofitMax}{107.55\xspace}
\newcommand{\BlurNWprofitMax}{5647.13\xspace}
\newcommand{\LooksRareFiloneCount}{57\xspace}
\newcommand{\LooksRareFiloneFrac}{0.01}
\newcommand{\LooksRareFiloneVol}{0.05}
\newcommand{\LooksRareFiltwoCount}{30171\xspace}
\newcommand{\LooksRareFiltwoFrac}{7.51}
\newcommand{\LooksRareFiltwoVol}{92.6}
\newcommand{\LooksRareFilthreeCount}{35770\xspace}
\newcommand{\LooksRareFilthreeFrac}{8.91}
\newcommand{\LooksRareFilthreeVol}{88.46}
\newcommand{\LooksRareFilfourCount}{147006\xspace}
\newcommand{\LooksRareFilfourFrac}{36.6}
\newcommand{\LooksRareFilfourVol}{80.19}
\newcommand{\LooksRareFilallfourCount}{163320\xspace}
\newcommand{\LooksRareFilallfourFrac}{40.66}
\newcommand{\LooksRareFilallfourVol}{95.68}
\newcommand{\LooksRareNWmedpriceMed}{0.0395\xspace}
\newcommand{\LooksRareWmedpriceMed}{0.0333\xspace}
\newcommand{\LooksRareWashtraderCount}{64465\xspace}
\newcommand{\LooksRareWashgroupCount}{6643\xspace}

\newcommand{\LooksRareNWprofitMedian}{0.01672\xspace}
\newcommand{\LooksRareWprofitMax}{69398.18\xspace}
\newcommand{\LooksRareNWprofitMax}{9861.77\xspace}
\newcommand{\OpenSeaFiloneCount}{2379\xspace}
\newcommand{\OpenSeaFiloneFrac}{0.01}
\newcommand{\OpenSeaFiloneVol}{0.0}
\newcommand{\OpenSeaFiltwoCount}{120910\xspace}
\newcommand{\OpenSeaFiltwoFrac}{0.56}
\newcommand{\OpenSeaFiltwoVol}{0.71}
\newcommand{\OpenSeaFilthreeCount}{1483348\xspace}
\newcommand{\OpenSeaFilthreeFrac}{6.89}
\newcommand{\OpenSeaFilthreeVol}{5.1}
\newcommand{\OpenSeaFilfourCount}{5574336\xspace}
\newcommand{\OpenSeaFilfourFrac}{25.91}
\newcommand{\OpenSeaFilfourVol}{22.43}
\newcommand{\OpenSeaFilallfourCount}{6717283\xspace}
\newcommand{\OpenSeaFilallfourFrac}{31.22}
\newcommand{\OpenSeaFilallfourVol}{26.39}
\newcommand{\OpenSeaNWmedpriceMed}{0.018\xspace}
\newcommand{\OpenSeaWmedpriceMed}{0.0192\xspace}
\newcommand{\OpenSeaWashtraderCount}{942034\xspace}
\newcommand{\OpenSeaWashgroupCount}{13194\xspace}

\newcommand{\OpenSeaWprofitMax}{2891.45\xspace}
\newcommand{\OpenSeaNWprofitMax}{124374.95\xspace}

\usepackage{changepage}

\title{Can AI Detect Wash Trading? Evidence from NFTs}

\author{
Brett Hemenway Falk, Gerry Tsoukalas, Niuniu Zhang\footnote{Hemenway Falk: University of Pennsylvania (\href{mailto:fbrett@cis.upenn.edu}{fbrett@cis.upenn.edu}), Tsoukalas: Boston University (\href{mailto:gerryt@bu.edu}{gerryt@bu.edu}), Zhang: University of California, Los Angeles (\href{mailto:niuniu.zhang.phd@anderson.ucla.edu}{niuniu.zhang.phd@anderson.ucla.edu})}
}

\date{February 22, 2025}

\begin{document}

\maketitle

\begin{abstract}
Existing studies on crypto wash trading often use indirect statistical methods or leaked private data, both with inherent limitations. This paper leverages public on-chain NFT data for a more direct and granular estimation. Analyzing three major exchanges, we find that $\sim$38\% (30--40\%) of trades and $\sim$60\% (25--95\%) of traded value likely involve manipulation, with significant variation across exchanges. This direct evidence enables a critical reassessment of existing indirect methods, identifying roundedness-based regressions à la \cite{cryptowashtrading} as most promising, though still error-prone in the NFT setting. To address this, we develop an AI-based estimator that integrates these regressions in a machine learning framework, significantly reducing both exchange- and trade-level estimation errors in NFT markets (and beyond).\\

{\bf Keywords:} AI, Boosting, Cryptocurrency, Decision Trees, Deep Learning, Non-Fungible Tokens (NFTs), Machine Learning, Market Manipulation, Wash Trading, Web3.
\end{abstract}


\newpage

\section{Introduction}

Web3 innovations hold the promise of decentralization, enhanced security, and greater transparency, fundamentally transforming various sectors. In particular, financial applications have arguably been at the forefront of this transformation since their inception. While these innovations have the potential to revolutionize the financial landscape, features such as account anonymization also introduce challenges, including novel methods of market manipulation. At the same time, they offer new opportunities for detecting and preventing such manipulations. For example, the on-chain nature of transactions produces transparent data that can be analyzed using statistical and machine learning tools. This paper explores the tension between the increased transparency offered by these technologies and the evolving strategies for market manipulation they enable.

In traditional stock and commodity markets, wash trading has been deemed illegal since 1936. This malpractice involves either a single actor or a coalition of actors engaging in self-directed trades to artificially manipulate market activity, aiming to exploit these distortions for profit. The SEC rigorously polices such activities in conventional financial markets; however, cryptocurrencies remain less stringently regulated. The anonymization of accounts mentioned above allows crypto traders to manipulate the appearance of market demand by easily creating and trading between multiple anonymous accounts. Accordingly, this subject is of considerable interest and debate to regulators, practitioners, and academics \citep{cryptowashtrading, decrypt2023wash, NFTWashProfit, GameofNFTs, NFTSuspicious}.

Addressing wash trading in cryptocurrencies presents a complex challenge. Approximately 83.3\% of crypto trades occur on private, centralized exchanges---such as Binance---that typically conduct off-chain transactions, thereby limiting transparency (\citealt{DEXsVol}). These platforms usually provide only rudimentary details (e.g., trading pair and size) while withholding trader identities, which complicates wash trading detection. Moreover, these exchanges may have little incentive to curb wash trading since inflated trading volumes can be beneficial. To overcome these challenges, current research leverages leaked insider data from hacks \citep{BitcoinWashTrading} and advanced statistical methods for pattern recognition \citep{cryptowashtrading}. The latter study estimates that up to 70\% of the traded value on certain exchanges may be attributed to wash trading---an alarming finding. While these approaches have advanced our understanding meaningfully, the lack of systematic access to timely public data continues to constrain their accuracy and their ability to capture live, ongoing market activity.

This paper seeks to bridge this gap by focusing on a specific segment of crypto markets: Non-Fungible Tokens (NFTs). These tokens are unique for their on-chain trading nature and inherent transaction transparency. NFTs are usually traded on decentralized exchanges, where every transaction is recorded on a public ledger, revealing comprehensive details including the identities (wallet addresses) of buyers and sellers. Such transparency facilitates a more direct analysis of trading patterns, better enabling the identification of practices like wash trading.  Consequently, NFTs provide a distinct advantage in data transparency and pattern identification compared to traditional centralized exchanges. In particular, NFT data can help answer several interrelated questions:

\subsubsection*{Research Questions}
First, can we leverage increased NFT transparency to establish a more direct estimation method for wash trading levels? 
Second, and perhaps more crucially, how do more direct estimation methods compare to indirect statistical approaches used in studies like \cite{cryptowashtrading}? Do the two methodologies align, and if discrepancies exist, is there a way to reconcile them? 

\subsubsection*{Data}
We concentrated our data collection on three of the top NFT marketplaces: LooksRare, Blur, and OpenSea. From these platforms, we manually amassed a comprehensive dataset encompassing all recorded transactions. Each transaction in our dataset is represented by a single row, detailing critical information such as the transaction hash, block number, NFT seller and buyer, the NFT collection, tokenId, price, and date of the transaction. Additionally, we scraped one degree upstream Ether transfers for all buyers and sellers from the sales dataset to identify possible collusion. This rich dataset offers an insightful view into the NFT trading landscape, encapsulating a diverse range of transactions across numerous collections. The depth of this data, covering thousands of NFT collections and a vast number of transactions with significant total value in ETH for both platforms, offers a comprehensive basis for our analysis of wash trading practices, allowing us to observe and decipher patterns and anomalies indicative of such activities. The details are discussed in Section~\ref{sec:data}.

\subsubsection*{Direct Estimation}
Using this dataset, we apply four distinct filters to directly estimate wash trading in NFT markets. Filter 1 flags the simplest case, where the buyer and seller use the same wallet, indicating a direct self-sale. Filter 2 detects back-and-forth trades—instances where buyer and seller identities are inverted in sequential transactions—suggesting coordinated trading between two accounts. Filter 3 identifies cases where the same buyer purchases the same NFT three or more times, capturing cyclical patterns unlikely to arise naturally. Filter 4 flags trades where both buyer and seller are funded by the same upstream account, signaling potential control by a single entity. Figure~\ref{fig:4filters} shows these filters graphically.  To reduce false positives in Filter 4, we exclude known addresses associated with smart contracts and exchanges. See Section~\ref{sec:direct_estimation}.

\subsubsection*{Indirect Estimation and Machine Learning}

Leveraging our on-chain data and direct estimation results, we assess the effectiveness of indirect estimation methods, i.e., methods that aim to classify the prevelance of wash trading, without identifying individual trades as wash trading. Building on \cite{cryptowashtrading}, we adopt their approach and distributional tests—examining deviations in trade-size roundedness, conformity to Benford’s Law, clustering around round numbers, and tail behavior—to detect potential market manipulation. We extend these methods in two key ways. First, we enhance their regression framework through hyperparameter optimization (see Section~\ref{sec:regressions}). Second, we embed the refined regression into a broader AI-driven filter that integrates machine learning techniques, including boosting and deep learning, with additional engineered features—such as trade price, round level, trading frequency, and inter-trade time intervals—to predict wash trading more accurately and at scale (see Section~\ref{sec:ml}).

\subsubsection*{Results}

Our results are presented as 11 key insights (I1–I11), detailed throughout the paper and summarized in Table~\ref{tab:filter_comparison} in the literature review (Section~\ref{sec:prior_work}). These findings highlight both the scale of NFT wash trading and the strengths and limitations of different detection methods.

First, in Section~\ref{sec:direct_estimation}, we apply our dataset and direct filters to estimate that approximately 38\% (30–40\%) of trades likely involve manipulation, with results remaining remarkably consistent across exchanges. However, these trades account for approximately 60\% (25–95\%) of traded value, reflecting substantial variation across marketplaces (I1). Variations aside, the overall average aligns with \cite{cryptowashtrading}, which estimates $\sim$70\% for \emph{fungible} tokens. Notably, collections implicated in wash trading tend to exhibit inflated prices (I2), though higher wash-trade activity does not systematically translate to greater profits for participants (I3).

Although NFTs differ structurally from cryptocurrencies, many of the incentives for wash trading remain similar. NFT marketplaces, like crypto exchanges, compete aggressively for liquidity and trader engagement, as higher reported trading volumes enhance platform rankings and attract investors. This dynamic is exemplified by Blur’s rapid rise past OpenSea through aggressive incentives. Additionally, NFTs introduce unique motivations for wash trading, including earning royalties per trade, benefiting from zero-fee trading models, and exploiting price opacity. On some platforms, we find wash trading volumes even exceed those observed in crypto exchanges, underscoring how these incentives can intensify market manipulation.

In Section~\ref{sec:regressions}, we adapt the roundedness regression method from \cite{cryptowashtrading}, to NFT markets. While this approach, which we term \emph{1-regression}, is effective in certain cases, it does not generalize seamlessly across all exchanges, leading to estimation discrepancies (I4). To address this, we refine the method by optimizing the roundedness threshold, calibrating it to each platform’s characteristics. This enhanced \emph{$\tau$-regression} reduces estimation errors to under 2\% on LooksRare (I5). 

However, residual discrepancies persist, motivating the introduction of our AI-based approach in Section~\ref{sec:ml}, which systematically reduces error rates below 2.15\% across all exchanges. The best models achieve AUC scores over 0.9 on NFT data (I6, I7). Notably, they perform even better outside  NFT markets (I8), particularly on the Mt. Gox dataset from \cite{BitcoinWashTrading},\footnote{We are thankful to the authors for facilitating access to the dataset.} where the AUC score exceeds 0.97, suggesting that our approach not only generalizes but may also be more effective in traditional fungible token markets.

Finally, in Section~\ref{sec:indirect_estimation}, we evaluate distributional tests—Benford’s Law, trade-size clustering, and power-tail distributions—which serve as coarse detectors of anomalous trading rather than precise estimators of wash trading volume. We find that both wash and legitimate NFT trades deviate from Benford’s Law, limiting its effectiveness (I9). Similarly, while trade-size clustering—based on the concentration of trades around round numbers—has been proposed as a wash-trading indicator, its effect appears muted in NFT markets (I10). Moreover, NFT trade-size distributions diverge significantly from the classical power-law behavior observed in other asset markets (I11). While these techniques provide useful benchmarks for flagging potential manipulation, their direct applicability to NFTs is limited.

Taken together, our contributions (I1–I10) establish a scalable framework for detecting and measuring wash trading at the individual trade level using public data. This framework not only benchmarks existing indirect methods but also introduces enhanced techniques—through optimized regression and machine learning—that bridge the gap between direct and indirect approaches, advancing our understanding of market manipulation in digital asset markets.

\section{Literature Review} \label{sec:prior_work}

Illicit market behaviors like wash trading on cryptocurrency exchanges and NFT marketplaces have drawn considerable academic attention. Early studies in traditional finance \cite{FraudTransaction, FinancialMarketWashTrading, ForensicEconomics, OddLotsMarketData} laid the foundation for detecting market manipulation, from identifying fraudulent transactions to analyzing odd-lot trades in equities. These works provide key insights into market scheming and detection methods, shaping the broader understanding of financial misconduct.

Among the most relevant studies to our work are \cite{cryptowashtrading} and \cite{BitcoinWashTrading}, which examine wash trading in cryptocurrency markets.

\cite{cryptowashtrading} introduce indirect statistical tests to detect fake transactions on cryptocurrency exchanges, estimating that over 70\% of traded value in unregulated markets stems from wash trading. Their approach provides valuable insights into the scale of manipulation but relies on statistical inference rather than direct observation and is not designed for trade-level detection.

Complementing this, \cite{BitcoinWashTrading} analyze Bitcoin wash trading using hacked Mt. Gox transaction logs, offering a rare opportunity for direct estimation \emph{on a centralized exchange} based on leaked trader identities. They define wash trades as self-to-self transactions\footnote{Their self-to-self trades align with our filter $f_1$. But the rest of our filters are specific to the on-chain nature of our data. For fungible tokens, filters $f_2$ and $f_3$ are likely to generate false positives.  For example, for filter $f_2$, a legitimate user is unlikely to sell a NFT to a buyer, then immediately buy the exact same NFT back from the same buyer, but a legitimate market maker might sell BTC, then buy BTC back moments later.  Our filter $f_3$, has similar issues when applied to \emph{fungible} tokens.  Finally, Filter $f_4$ cannot be applied to their dataset because it only includes Mt. Gox trading data, which does not have information on how to link wallets to their original funding source the way we can using on-chain data.} to establish a lower bound and evaluate several other direct and indirect estimation methods, including those from \cite{cryptowashtrading}, but find limited applicability in their dataset. For example, their direct filters (before augmentation with insider reports) estimate around 3\% of wash trading activity on Bitcoin, compared to an average of 70\% in \cite{cryptowashtrading}.

Our study builds on these seminal contributions while addressing some inherent limitations. The methods in \cite{cryptowashtrading} are designed for aggregate-level estimation, whereas we focus on trade-level predictions. Similarly, \cite{BitcoinWashTrading} rely on leaked data and insider reports, providing a static snapshot of a single exchange. In contrast, we leverage public NFT data and apply a broader set of direct filters, allowing for a scalable and adaptable approach across multiple marketplaces and collections. This direct estimation framework then enables us to introduce both an optimized log-log regression method and a machine-learning-based estimator, reducing estimation errors between direct and indirect methods by more than an order of magnitude.

Beyond these key differences, our work also diverges from \cite{cryptowashtrading} and \cite{BitcoinWashTrading} in several other dimensions, summarized in Table~\ref{tab:filter_comparison}.

\begin{table}[h]
\centering
\resizebox{.98\textwidth}{!}{%
\begin{tabular}{@{}lccc@{}}
\toprule
& \textbf{Our Paper} & 
\textbf{Aloosh \& Li (2024)} & \textbf{Cong et al. (2023)} \\ 
\midrule
\textbf{Dataset}\\
Assets & NFTs & Bitcoin  & Crypto\\ 
Multiple tokens/currencies? & $\checkmark$ & $\times$ & $\checkmark$\\
Multiple exchanges? & $\checkmark$ & $\times$ & $\checkmark$\\
Public On-Chain Data? & $\checkmark$ & $\times$ & $\times$\\
Transaction-level estimators? & $\checkmark$ & $\checkmark$ & $\times$\\
All methods apply without leaked info? & $\checkmark$ & $\times$ & $\checkmark$\\
\midrule
\textbf{Efficacy of Distributional Filters} \\
Benford (Section \ref{sec:benford_comparison}) & $\times$  & $\checkmark$  & $\checkmark$   \\ 
Trade Size Clustering (Section \ref{sec:clustering_comparison})& $\checkmark$ & $\times$ & $\checkmark$ \\ 
Power Tail  (Section \ref{sec:tail_comparison})    & $\times$ &  $\times$   & $\checkmark$   \\ 
\midrule
\multicolumn{4}{@{}l}{\textbf{Avg. Wash Value \% - Direct}} \\
\textcolor{black}{Our $\cup f_i$} filters (Section~\ref{sec:direct_estimation}) & 60.08\% & $\times$  & $\times$ \\ 
Aloosh \& Li filters & $\times$ & 34.39\% & $\times$\\
\midrule
\multicolumn{4}{@{}l}{\bf Avg. Wash Value Estimation \% - Indirect}\\
1-Regression (Section~\ref{sec:modreg}) & 46.98\% & $3.30\%$  & 70.85\% \\ 
\textcolor{black}{$\tau$-Regression}  (Section~\ref{sec:taureg})  & 52.75\% & $\times$  & $\times$ \\ 
\textcolor{black}{Machine Learning (Section~\ref{sec:ml}})  & 59.54\% & $\times$  & $\times$ \\ 
\midrule
\multicolumn{4}{@{}l}{\textbf{Avg. Wash Value Estimation \% Error}} \\
1-Regression (Section~\ref{sec:modreg}) & 13.09\% & 31.09\% & $\times$ \\ 
\textcolor{black}{$\tau$-Regression}  (Section~\ref{sec:taureg}) & 7.33\% & $\times$ & $\times$ \\ 
\textcolor{black}{Machine/Deep Learning (Section~\ref{sec:ml}})  & 0.54\% & $\times$  & $\times$ \\ 
\bottomrule
\end{tabular}%
}
\caption{Comparison to seminal papers.}
\label{tab:filter_comparison}
\end{table}

Beyond these foundational papers, several other studies explore wash trading and market manipulation from different angles.

\cite{NFTWashProfit} analyze the profitability of NFT wash trading, while \cite{GameofNFTs} examine how traders exploit token reward systems on NFT platforms. \cite{NFTSuspicious} apply a graph-theoretic approach to quantify cyclic and non-cyclic wash trades in decentralized Ethereum NFT markets. \cite{TradingVolumeManipulation} investigate the link between wash trading and platform competition among centralized crypto exchanges.

Other studies focus on specific market segments. \cite{DarkSideofNFT} analyze wash trading within OpenSea’s popular NFT collections but on a more limited dataset than our multi-platform study. \cite{MarketManipulation} explore the economic motivations behind wash and insider trading in Ethereum NFTs but do not benchmark detection methods. \cite{TaxLoss} examine tax-loss harvesting in cryptocurrencies, highlighting investor-driven incentives distinct from exchange-driven manipulation. These studies use different methodologies, and their wash-trading estimates often diverge significantly from those in \cite{cryptowashtrading}, whereas ours remain consistent.

Broader NFT market research also provides relevant context. \cite{NFTBubble} examine how behavioral biases, such as selection neglect, influence NFT markets during speculative bubbles, while \cite{NFTHerding} analyze herding behavior in “blue-chip” NFTs, shedding light on price formation and liquidity patterns.

While these studies offer valuable insights, none systematically compare direct and indirect detection methods across different markets. Our work fills this gap by evaluating the strengths and limitations of each approach, enabling more precise wash-trading detection and laying the foundation for new methodologies tailored to diverse market structures.\footnote{Beyond academic research, practitioners also see value in using direct filters to estimate wash trading, e.g., this blog post by \cite{washtrading}. This said, we are not aware of any practitioner publication that systematically analyzes the relationship between filter-based estimation and indirect statistical methods, as we do.}

\section{Data} \label{sec:data}

Our study focuses on the NFT marketplaces OpenSea, LooksRare and Blur, selected due to their significant and different roles in the NFT market. 

As of May 2024, OpenSea held the number one position by all-time volume with \$36.9 billion, followed by Blur with \$10.54 billion, and LooksRare in third place with \$4.86 billion.\footnote{\url{https://dappradar.com/rankings/nft/marketplaces?range=all}} This selection is particularly relevant for analyzing different aspects of NFT wash trading: LooksRare is known for its problematic token reward system, which has been criticized for paving the way to severe wash trading, whereas a reward system is absent in OpenSea \citep{GameofNFTs}. Blur's program, more focused on user loyalty and platform engagement, offers a contrasting perspective \citep{metadigest2023blur}.

We obtained transaction data of LooksRare, Blur, and OpenSea through the Alchemy \href{https://docs.alchemy.com/reference/getnftsales}{getNFTSales} API. For LooksRare, our data covers all transactions from \LooksRareTimeMin to \LooksRareTimeMax. For Blur, our data covers all transactions from \BlurTimeMin to \BlurTimeMax. For OpenSea, our data covers all transactions from \OpenSeaTimeMin to \OpenSeaTimeMax.
In these datasets, each transaction is represented by a single row, with the schema detailed as follows:
\begin{itemize}
  \item \textbf{Transaction hash:} A unique identifier for the transaction on the blockchain.
  \item \textbf{Block number:} The specific block on the blockchain where the transaction is recorded.
  \item \textbf{NFT seller:} The address of the individual or entity selling the NFT.
  \item \textbf{NFT buyer:} The address of the individual or entity purchasing the NFT.
  \item \textbf{NFT collection:} The specific collection to which the NFT belongs.
  \item \textbf{NFT tokenId:} A unique identifier for the specific NFT within its collection.
  \item \textbf{Price:} The sale price of the NFT, denominated in ETH. 
  \item \textbf{Date:} The date and time when the transaction occurred.
\end{itemize}

The datasets from LooksRare, Blur, and OpenSea exchanges are substantial. The Blur exchange dataset occupies 1.4 GB and comprises \BlurTradeCounts transactions, the LooksRare dataset is 194.2 MB with \LooksRareTradeCounts transactions, and the OpenSea dataset is 10.6 GB with \OpenSeaTradeCounts transactions.  For data cleaning, the direct outputs from Alchemy for Blur and LooksRare are ready as is. However, for OpenSea, we combined transactions from both the Seaport and Wyvern protocols and excluded the 0.39\% of transactions not conducted in ETH-type currencies (e.g., ETH, wETH) to ensure consistency in our analysis.

To identify potential collusions between NFT traders (buyer and seller), we collected one degree upstream transfers of all NFT traders per platform using Alchemy's \href{https://docs.alchemy.com/reference/alchemy-getassettransfers}{getAssetTransfers} API. For each marketplace, we specified the to address as this marketplace's unique trader, set fromBlock to be well before the marketplace's contract deployment date, and toBlock to be well after the last transaction in the marketplace's sales data. To err on the conservative side, we limited the transfers to be of ``external" type, i.e., Ether transfers from Externally Owned Accounts (EOAs).\footnote{\href{https://docs.alchemy.com/reference/transfers-api-quickstart\#types-of-transfers}{https://docs.alchemy.com/reference/transfers-api-quickstart\#types-of-transfers}} 

The resulting upstream dataset per marketplace includes schemas that contain ``to" (unique buyers and sellers from the marketplace sales dataset) and all the ``from" EOAs who sent Ether, and Ether only, to these traders. The upstream data sizes are 1.6 GB for Blur, 2.3 GB for LooksRare, and 14 GB for OpenSea, with a total of 6,961,436 Ether transfers for Blur, 10,100,725 Ether transfers for LooksRare, and 56,845,125 Ether transfers for OpenSea.

In total, we identified \LooksRareCollectionCount NFT collections on LooksRare, \BlurCollectionCount NFT collections on Blur, and \OpenSeaCollectionCount NFT collections on OpenSea. Figure~\ref{fig:LR_total_sales} shows a histogram of total sale value per collection on LooksRare, Figure~\ref{fig:Blur_total_sales} shows a histogram of total sales value per collection on Blur, and Figure~\ref{fig:OpenSea_total_sales} shows a histogram of total sales value per collection on OpenSea. The figures exclude any outlier collections that amassed total sale value above the 99th percentile - \LooksRareSalesNinetynine ETH on LooksRare, \BlurSalesNinetynine ETH on Blur, and \OpenSeaSalesNinetynine ETH on OpenSea respectively. Each bin of histograms represents an even $1/100$ intervals from 0 to 99th percentile of total sale value.

\subsection*{Data Visualization and Summary Statistics}

\begin{figure}[H]
\begin{adjustwidth}{-1cm}{-1cm} 
\begin{center}
\begin{subfigure}[t]{0.35\textwidth}
  \centering
  \includegraphics[width=\linewidth]{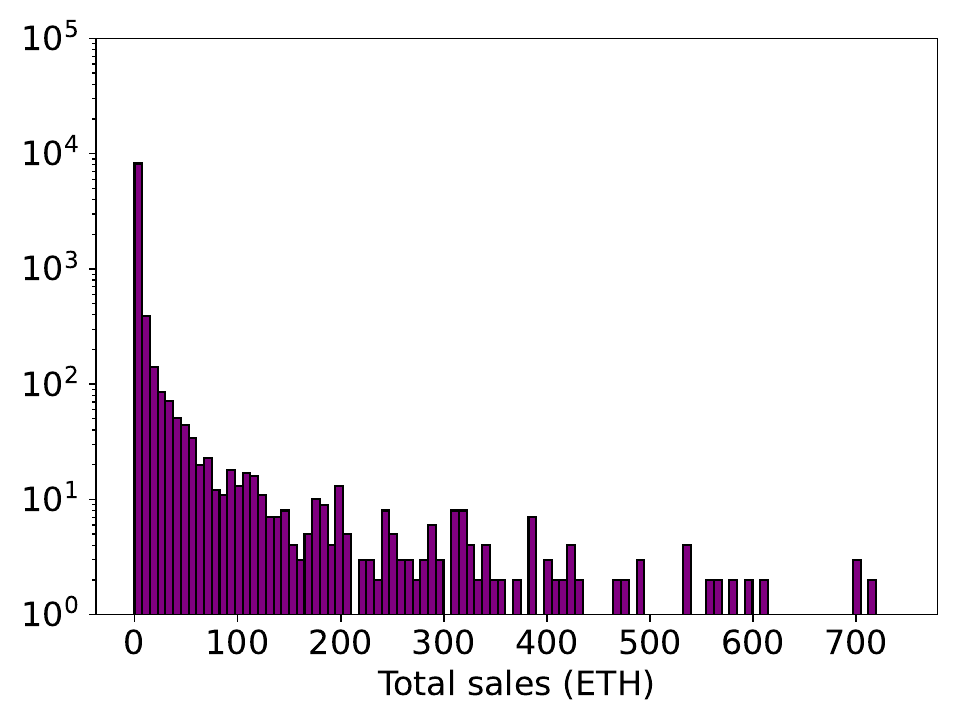}
  \caption{LooksRare}
  \label{fig:LR_total_sales}
\end{subfigure}%
\hspace{-0.1cm}%
\begin{subfigure}[t]{0.35\textwidth}
  \centering
  \includegraphics[width=\linewidth]{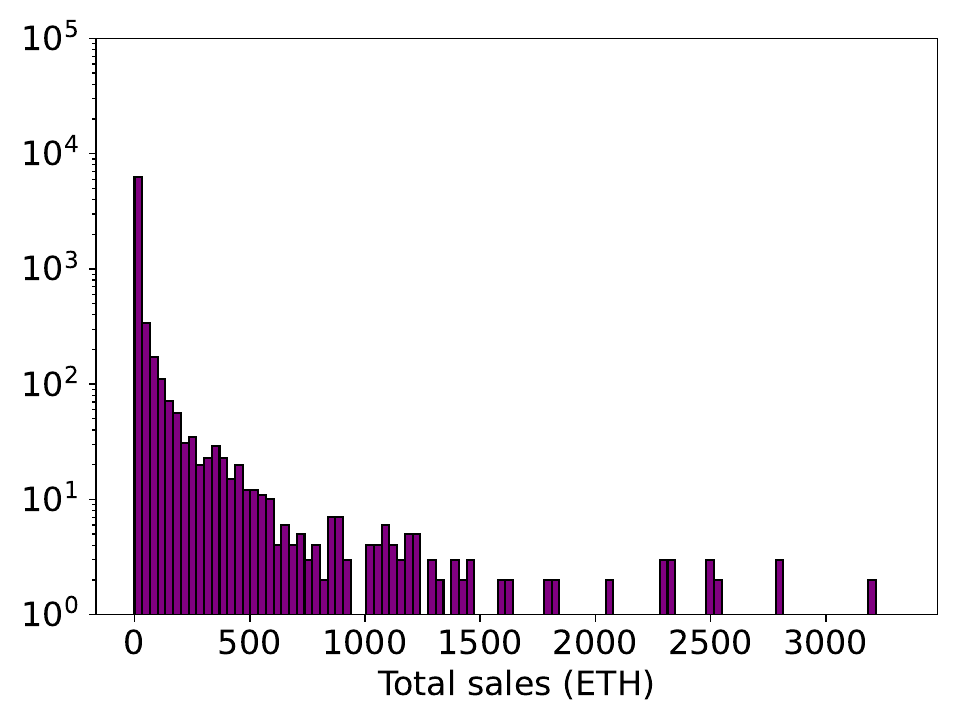}
  \caption{Blur}
  \label{fig:Blur_total_sales}
\end{subfigure}%
\hspace{-0.1cm}%
\begin{subfigure}[t]{0.35\textwidth}
  \centering
  \includegraphics[width=\linewidth]{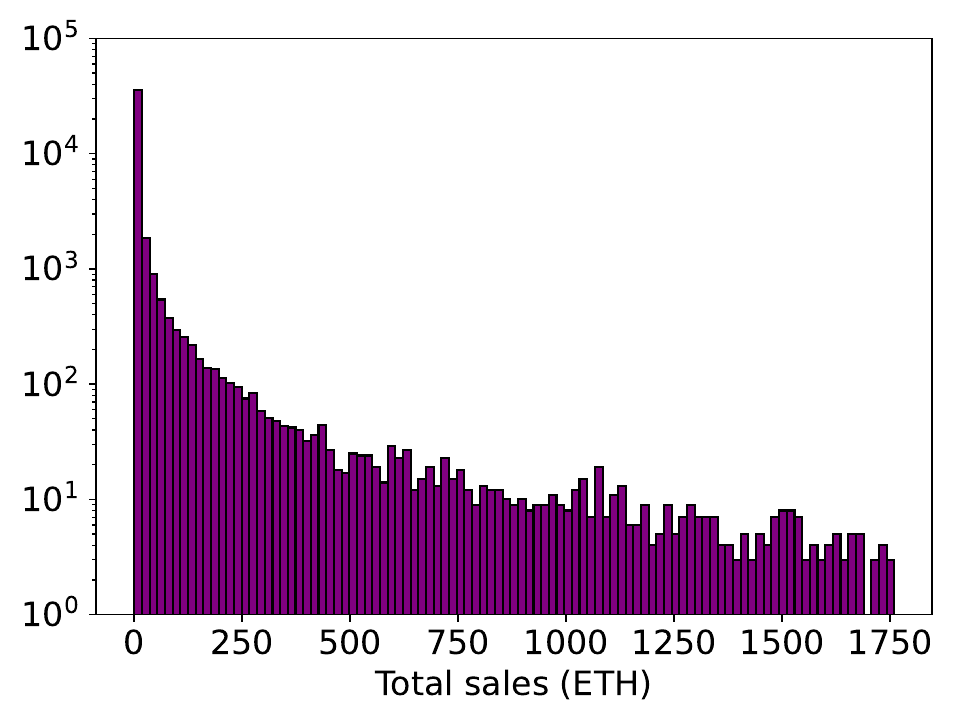}
  \caption{OpenSea}
  \label{fig:OpenSea_total_sales}
\end{subfigure}
\end{center}
\end{adjustwidth}
\caption{Distribution of total value traded.}
\label{fig:comparison_LR_Blur_OpenSea}
\end{figure}

Figure~\ref{fig:comparison_LR_Blur_OpenSea} illustrates the distribution of trade volumes per NFT collection on each platform. We exclude the top 1\% of collections to prevent scale distortion from exceptionally high transaction prices, thus offering a clearer comparison across the broader range of collections.

The histograms for all three platforms display a distribution that is heavily skewed to the left, indicating that a majority of NFT collections fall into the lower sales value category. This suggests that while there are a large number of NFT collections available on both platforms, the bulk of these collections are what might be termed ``small fish,'' with only a few reaching higher sales value. LooksRare and Blur show sparser distributions towards the right, while OpenSea appears ``smoother" due to its significantly higher trade count of \OpenSeaTradeCounts compared to \BlurTradeCounts for Blur and \LooksRareTradeCounts for LooksRare. This concentration in the lower end of the sales spectrum illustrates the long-tail nature of NFT collections in these marketplaces.

A technical note: transactions on LooksRare involve trading NFT for Ether. OpenSea includes sales done in Ether, Wrapped Ether, and other ERC20 tokens. Blur involves a mixture of Ether, Wrapped Ether (\href{https://etherscan.io/address/0xC02aaA39b223FE8D0A0e5C4F27eAD9083C756Cc2}{wETH}), as well as their own \href{https://etherscan.io/address/0x0000000000a39bb272e79075ade125fd351887ac}{Blur Pool token}.\footnote{Blur implements its own token to reduce gas costs when interacting with the Blur exchange contract.} Like wrapped Ether, Blur pool tokens can be minted by depositing ETH, and redeemed for ETH in the pool. Since all three tokens (ETH, wETH, and Blur Pool Tokens) trade at the same valuation, we can treat all payments as being in Ether. Our LooksRare data set has \LooksRareTradeCounts transactions that amounts to a total value of \LooksRareTotalVol ETH. As for the Blur data set, it has \BlurTradeCounts transactions that amounts to a total value of \BlurTotalVol ETH. Our OpenSea dataset, after filtering to include only sales done in Ether, wETH, and Blur Pool Tokens, resulted in \OpenSeaTradeCounts transactions (99.61\% of the raw OpenSea sales data) amounting to \OpenSeaTotalVol ETH.

\section{Direct Estimation}\label{sec:direct_estimation}

In developing our direct estimator for wash trading, we introduce four filters specifically tailored to accommodate observables from on-chain NFT data. We discuss each filter in detail next.

\subsection{Wash Trading Filters \label{sec:filters}}

\begin{figure}[H]
    \centering
    \tikzstyle{user}=[circle,minimum width=5mm,draw]
    \begin{tikzpicture}
        \node (F1) at (0,0) [user] {};
        \draw [->,out=40,in=320,looseness=10] (F1) to (F1);
        \node (T1) at ([yshift=-1cm]F1) {Filter $f_1$: Self trades};

        \coordinate (F2) at ([xshift=5cm]F1);
         \foreach \x/\y in {A/0,B/180} {
             \node (F2\x) at ($(F2)+(\y:10mm)$) [user] {};
        }
        \draw [->,out=150,in=30] (F2A) to (F2B);
        \draw [->,out=330,in=210] (F2B) to (F2A);
        \node at (T1-|F2) {Filter $f_2$: Back-and-forth trades};

        \coordinate (F3) at ([yshift=-3.5cm]F1);
         \foreach \x/\y in {A/0,B/120,C/240} {
             \node (F3\x) at ($(F3)+(\y:10mm)$) [user] {};
        }
        \draw [->,out=90,in=30] (F3A) to (F3B);
        \draw [->,out=85,in=35] (F3A) to (F3B);
        \draw [->,out=95,in=25] (F3A) to (F3B);

        \draw [->,out=210,in=150] (F3B) to (F3C);
        \draw [->,out=215,in=145] (F3B) to (F3C);
        \draw [->,out=205,in=155] (F3B) to (F3C);

        \draw [->,out=330,in=270] (F3C) to (F3A);
        \draw [->,out=335,in=265] (F3C) to (F3A);
        \draw [->,out=325,in=275] (F3C) to (F3A);

        \node at ([yshift=-1.5cm]F3) {Filter $f_3$: 3 times around};

        \coordinate (F4) at ([yshift=-3.5cm]F2);
         \foreach \x/\y in {A/-30,B/90,C/210} {
             \node (F4\x) at ($(F4)+(\y:10mm)$) [user] {};
        }
        \draw [->,out=330,in=120] (F4B) to node [right] {\tiny ETH} (F4A);
        \draw [->,out=210,in=60] (F4B) to node [left] {\tiny ETH} (F4C);
        \draw [->,out=0,in=180] (F4C) to node [above] {\tiny NFT} (F4A);

        \node at ([yshift=-1.5cm]F4) {Filter $f_4$: Common funder};
    \end{tikzpicture}
    \caption{The four direct wash-trading filters.}
    \label{fig:4filters}
\end{figure}
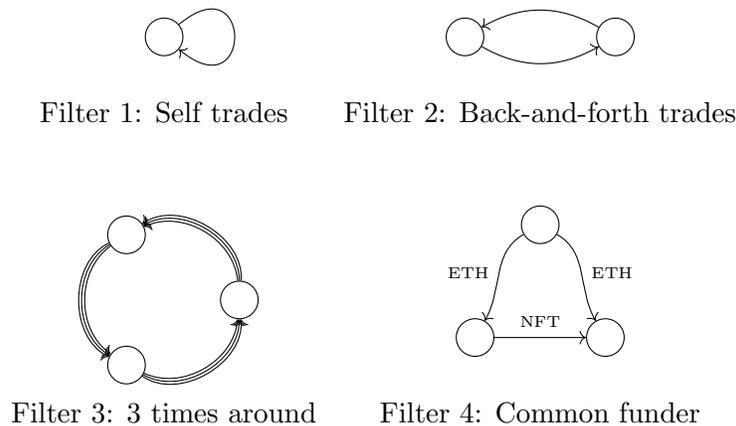

Figure~\ref{fig:4filters} provides a visual representation of the four filters. 

Filter $f_1$ flags trades where the buyer and seller account is one and the same -- this is the most na\"{i}ve form of wash-trading -- literally selling the NFT to oneself. 

Filter $f_2$ detects back and forth trades, meaning, if buyer and seller are inverted for the same exact NFT, then $f_2$ will be triggered. This is intended to flag slightly more sophisticated wash trading activity, where the wash trader has created two accounts and sells the same NFT back and forth between these two accounts. It gets triggered even when there is just one instance of inversion between buyer and seller identities for the exact same NFT. This means if an NFT is sold from Address $A$ to Address $B$, and then subsequently sold back from Address $B$ to Address $A$,  $f_2$ would identify this as a potential wash trade.

Filter $f_3$ flags trades where the same buyer purchased the same NFT, three or more times.  Imagine a slightly more sophisticated wash trader, who generates multiple (more than two) accounts, and trades a single NFT between these accounts in a cycle.  Unlike $f_2$, which flags direct back-and-forth trades between two accounts, $f_3$ identifies circular trading patterns involving three or more accounts. Filter $f_3$ will flag this activity as wash trading, if the NFT passes around the cycle at least three times. For example, if an NFT is first sold to Buyer $A$, who then sells it to Buyer $B$, and Buyer $B$ subsequently sells it to Buyer $C$, who then completes the loop by selling it back to Buyer $A$, $f_3$ would flag the activity when Buyer $A$ acquires the NFT for the third time. This cycle could involve numerous parties and is designed to detect more complex wash trading schemes where a single entity might be operating multiple accounts to simulate a closed loop of trades.

Finally, Filter $f_4$ detects transactions with a common upstream wallet that has funded both the buyer and seller, indicating potential control over both sides of the trade. In our dataset, we analyze the transfer history up to one degree upstream for both buyers and sellers. This means, for example, if Wallet $A$ sells an NFT to Wallet $B$, the Filter checks if there is a Wallet $C$ that previously transferred funds (Ether only) to both Wallet $A$ and Wallet $B$. The presence of such a common upstream wallet suggests a single entity may be orchestrating the trade from behind the scenes, using different wallets to create the illusion of a genuine transaction.

\subsection{Mitigating False Positives}

We believe our Filters 1-3 are extremely conservative, and unlikely to classify organic behavior as wash trading.  By focusing only on NFT \emph{sales}, rather than NFT \emph{transfers}, we immediately exclude all activity where a single user simply \emph{transfers} an NFT between different wallets, e.g. to increase anonymity, or reduce risk.  Second, Filters 1-3, only catch behavior where a user engages with the NFT marketplace (and pays trading fees to the marketplace on top of the Ethereum gas costs), and the \emph{same NFT} ends up in exactly the \emph{same wallet}.

When implementing these filters, we have to make a decision about granularity.  For example, in $f_2$, when we say ``the same'' NFT was traded back and forth between, does ``the same'' NFT mean the same NFT contract (e.g. Bored Apes) or the same exact NFT (e.g. \href{https://ipfs.io/ipfs/QmeSjSinHpPnmXmspMjwiXyN6zS4E9zccariGR3jxcaWtq/3401}{Bored Ape \#3401}).  We can ask a similar question for $f_3$ as well.  For $f_2$, we consider a typical wash trader would buy and sell from the same NFT collections (e.g. Bored Apes) back and forth, but not necessarily the same token. Thus, we set $f_2$ at the collection level. For $f_3$, since it is reasonable for a legitimate trader to buy multiple tokens from the same collection, we apply $f_3$ at the individual token level, (e.g. \href{https://ipfs.io/ipfs/QmeSjSinHpPnmXmspMjwiXyN6zS4E9zccariGR3jxcaWtq/3401}{Bored Ape \#3401}), rather than at the collection level. This ensures that frequent purchases within a collection are not automatically classified as wash trading, while repeated trades of the same token remain a strong indicator of wash activity.

To mitigate potential false positives in $f_4$, which identifies common upstream wallets, we exclude \emph{all} smart contract addresses and EOAs of centralized exchanges (CEXs). Legitimate traders often interact with the same smart contracts. For instance, if both buyer and seller minted Wrapped Ether (wETH), they would have received a ``payment" from the wETH smart contract. Similarly, if both parties sold NFTs on OpenSea before transacting on LooksRare, they could mistakenly be flagged due to payments from OpenSea's contract.

Furthermore, since some CEXs also use EOAs, such as Binance deposit addresses, we also exclude transactions where the upstream EOA is tagged as belonging to a CEX. While this method yields a conservative estimate by excluding other type of transfers, e.g., ERC20 tokens transfers, it reflects common wash trading practices where Ether is typically used to cover gas and transaction costs. Applying these measures, we ensure that only direct, user-controlled funding sources are considered.

\subsection{Direct Estimation Results}

Table~\ref{tab:CombinedWashTrades} shows our direct estimation results after applying each filter individually, and then combining them ($\cup f_i$).

\begin{table}[h]
\centering
\begin{tabular}{c l c c c}
\toprule
{}
 & {}
 & \textbf{Wash Trades} 
 & \textbf{Wash Value} 
 & {} \\ 
\textbf{Filter} 
 & \textbf{Marketplace} 
 & \textbf{(\% of transactions)} 
 & \textbf{(\% of value traded)} 
 & \textbf{Count} \\  
\midrule
\multirow{3}{*}{$f_1$} 
 & LooksRare & \LooksRareFiloneFrac\% & \LooksRareFiloneVol\% & \LooksRareFiloneCount \\
 & Blur      & \BlurFiloneFrac\%      & \BlurFiloneVol\%      & \BlurFiloneCount      \\
 & OpenSea   & \OpenSeaFiloneFrac\%   & \OpenSeaFiloneVol\%   & \OpenSeaFiloneCount   \\
\midrule
\multirow{3}{*}{$f_2$} 
 & LooksRare & \LooksRareFiltwoFrac\% & \LooksRareFiltwoVol\% & \LooksRareFiltwoCount \\
 & Blur      & \BlurFiltwoFrac\%      & \BlurFiltwoVol\%      & \BlurFiltwoCount      \\
 & OpenSea   & \OpenSeaFiltwoFrac\%   & \OpenSeaFiltwoVol\%   & \OpenSeaFiltwoCount   \\
\midrule
\multirow{3}{*}{$f_3$} 
 & LooksRare & \LooksRareFilthreeFrac\% & \LooksRareFilthreeVol\% & \LooksRareFilthreeCount \\
 & Blur      & \BlurFilthreeFrac\%      & \BlurFilthreeVol\%      & \BlurFilthreeCount      \\
 & OpenSea   & \OpenSeaFilthreeFrac\%   & \OpenSeaFilthreeVol\%   & \OpenSeaFilthreeCount   \\
\midrule
\multirow{3}{*}{$f_4$} 
 & LooksRare & \LooksRareFilfourFrac\% & \LooksRareFilfourVol\% & \LooksRareFilfourCount \\
 & Blur      & \BlurFilfourFrac\%      & \BlurFilfourVol\%      & \BlurFilfourCount      \\
 & OpenSea   & \OpenSeaFilfourFrac\%   & \OpenSeaFilfourVol\%   & \OpenSeaFilfourCount   \\
\midrule
\multirow{4}{*}{$\cup f_i$}
 & LooksRare & \LooksRareFilallfourFrac\% & \LooksRareFilallfourVol\% & \LooksRareFilallfourCount \\
 & Blur      & \BlurFilallfourFrac\%      & \BlurFilallfourVol\%      & \BlurFilallfourCount      \\
 & OpenSea   & \OpenSeaFilallfourFrac\%   & \OpenSeaFilallfourVol\%   & \OpenSeaFilallfourCount   \\
 \cmidrule{2-5}
 & Average   & 37.58\%   & 60.08\%   & 2770713   \\ 
\bottomrule
\end{tabular}
\caption{Direct estimation of wash trades by filter and marketplace.}
\label{tab:CombinedWashTrades}
\end{table}

Wash trades, as a percentage of total trades, appear quite stable across exchanges -- we estimate between 30-40\% across. On the other hand, there is significant heterogeneity (26\% - 95\%) in terms of wash trade values (often referred to as traded volume in the literature), that is, when taking into account not just the number of trades categorized as wash, but also their size and price. 

On LooksRare, Filter 1 only flagged \LooksRareFiloneCount trades, whereas Filter 4 flagged \LooksRareFilfourCount trades as wash trades. In total, filters 1 through 4 collectively identified \LooksRareFilallfourCount wash trades, which is not simply the sum of the individual filters' results. This is because a single transaction may meet the criteria of multiple filters, and the last line in the table represents any transaction flagged by at least one of these filters (logical OR). Overall, we found \LooksRareFilallfourFrac\% of wash trades on LooksRare corresponding to an astouding \LooksRareFilallfourVol\% of the trade value in ETH (in other words, trades identified as wash, were larger in size on average
). 

On Blur, Filter 1 flagged the least, \BlurFiloneCount trades, while Filter 4 flagged the most, \BlurFilfourCount trades. Here, the combination of filters 1 through 4 caught \BlurFilallfourCount trades, with \BlurFilallfourFrac\% of transactions accounting for \BlurFilallfourVol\% of the value.

Finally, on OpenSea, Filter 1 identified \OpenSeaFiloneCount trades as wash trades, while Filter 4 flagged \OpenSeaFilfourCount trades. Collectively, filters 1 through 4 flagged \OpenSeaFilallfourCount trades, where \OpenSeaFilallfourFrac\% of the flagged wash trades contributed \OpenSeaFilallfourVol\% of the total value. 

The discrepancy in wash trade value between LooksRare (\LooksRareFilallfourVol\%), Blur (\BlurFilallfourVol\%), and OpenSea (\OpenSeaFilallfourVol\%), despite their similar fraction of wash trades (\LooksRareFilallfourFrac\%, \BlurFilallfourFrac\%, and \OpenSeaFilallfourFrac\%, respectively), can be partly attributed to findings from \cite{GameofNFTs}. Their study suggests that LooksRare, while processing fewer transactions, focuses on higher-value NFTs, which likely contributes to the observed difference in wash trade value

\subsection{Impact of Wash Trading on NFT Prices and Trader Profits}

We can leverage our direct estimator to investigate two additional questions that may be of interest: does wash trading influence NFT prices and trader profits? Although our analysis reveals correlations, we do not assert causation. Given that our primary focus is on comparing direct and indirect estimation methods, these findings are presented as ancillary observations to inform future research.

\subsubsection*{Wash trading correlates with higher NFT prices}

\begin{figure}[ht]
\begin{adjustwidth}{-1cm}{-1cm} 
\begin{center}
  \begin{subfigure}[t]{0.35\textwidth}
    \centering
    \includegraphics[width=\linewidth]{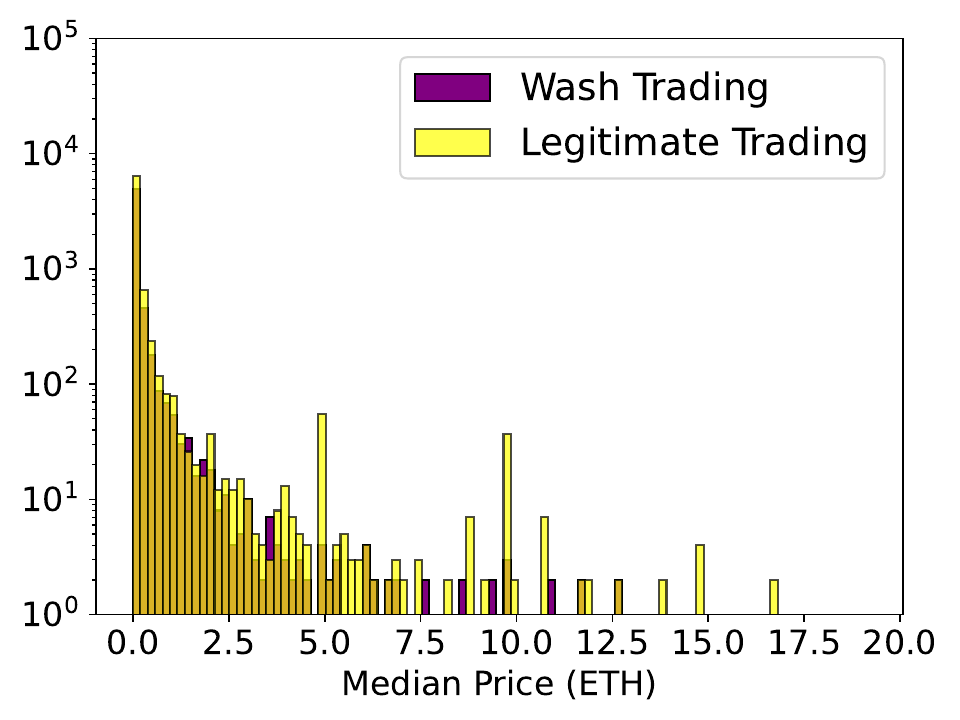}
    \caption{LooksRare}
    \label{fig:LooksRare_median_price}
  \end{subfigure}%
  \hspace{-0.1cm}%
  \begin{subfigure}[t]{0.35\textwidth}
    \centering
    \includegraphics[width=\linewidth]{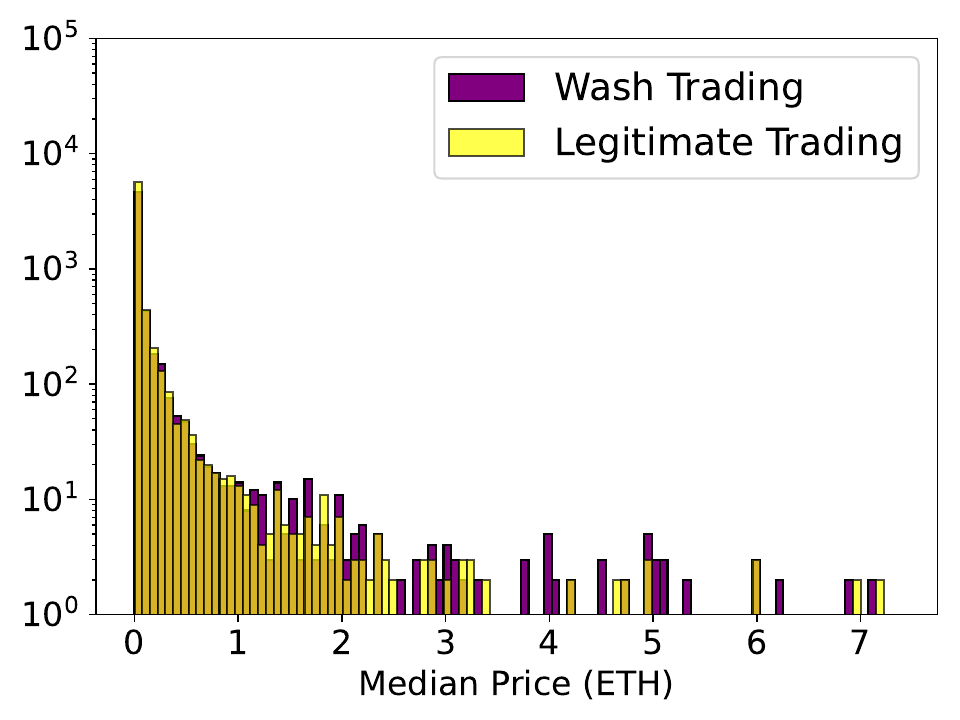}
    \caption{Blur}
    \label{fig:Blur_median_price}
  \end{subfigure}%
  \hspace{-0.1cm}%
  \begin{subfigure}[t]{0.35\textwidth}
    \centering
    \includegraphics[width=\linewidth]{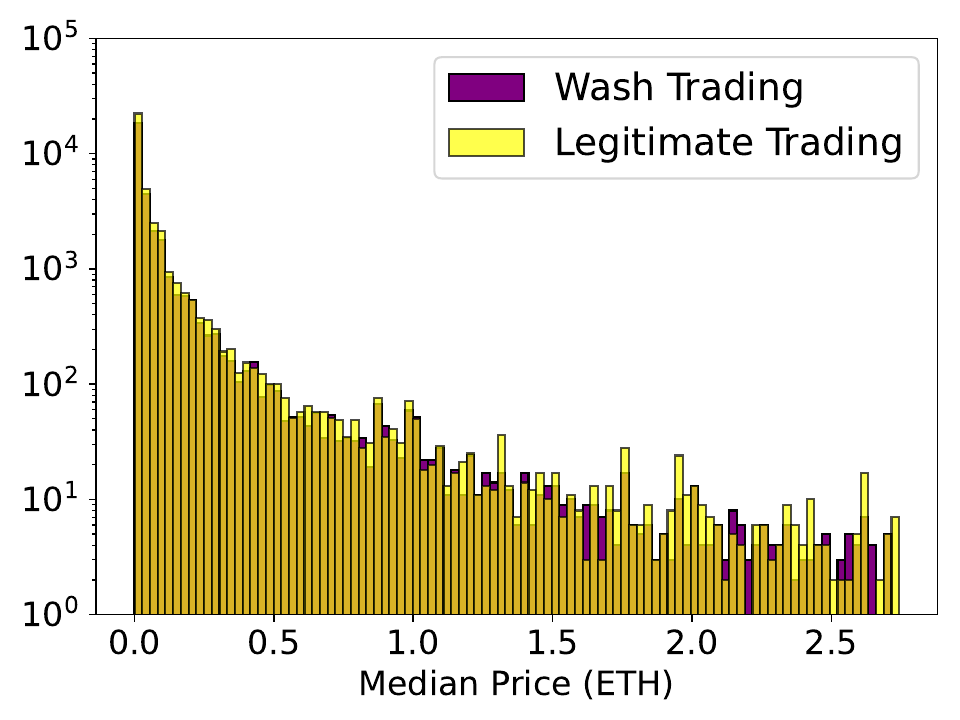}
    \caption{OpenSea}
    \label{fig:OpenSea_median_price}
  \end{subfigure}
\end{center}
\end{adjustwidth}
\caption{Distribution of median transaction prices.}
\label{fig:median_prices_comparison}
\end{figure}

To examine the impact of wash trading on NFT prices, we categorized NFT collections into two groups: those involved in wash trading and those that were not. We then calculated the median transaction price for each collection, excluding the top 1\% of sales to reduce the influence of outliers. Figure \ref{fig:median_prices_comparison} presents the resulting histograms on a logarithmic scale, where purple bars represent wash-traded collections, yellow bars indicate legitimate trading, and overlapping areas appear as a darker yellow.

Our analysis reveals significant differences across exchanges. The median transaction price is \LooksRareMedpriceNinetynine ETH on LooksRare, \BlurMedpriceNinetynine ETH on Blur, and \OpenSeaMedpriceNinetynine ETH on OpenSea, with the highest-priced collections reaching \LooksRareMedpriceMax ETH, \BlurMedpriceMax ETH, and \OpenSeaMedpriceMax ETH, respectively.

Wash-traded collections exhibit higher median sale prices  on Blur and OpenSea, but the opposite is observed on LooksRare. Specifically, the median sale prices for wash-traded collections are \LooksRareWmedpriceMed ETH on LooksRare, \BlurWmedpriceMed ETH on Blur, and \OpenSeaWmedpriceMed ETH on OpenSea, compared to \LooksRareNWmedpriceMed ETH, \BlurNWmedpriceMed ETH, and \OpenSeaNWmedpriceMed ETH for collections without wash trading. These findings suggest that while wash trading may artificially inflate the perceived value for some NFT collections, it is not  universal.

\subsubsection*{Wash trading does not correlate with increased wash trader profits}
Given wash trading seems to lead to increased median prices, it naturally follows to ask whether this activity actually increases short-term wash trader profits. To help answer this question, we grouped wash trading accounts based on their connectivity through trades. In particular, as wash trading usually involves more than one wallet, we group buyer and seller as one entity if their transaction is flagged as a wash trade. To illustrate, consider the following toy example involving 3 traders and 2 transactions. Suppose there are traders $A, B$, and $C$, in which the transaction from $A$ to $B$ is flagged as wash trade and the one from $B$ to $C$ is not.  

\begin{center}
\begin{tikzcd}
A \arrow[rr, "\text{Wash trade}"] &   & B \arrow[ldd, "\text{Legitimate trade}" description] &       & {\fbox{$A,B$}} \arrow[dd, "\text{Legitimate trade}" description] \\
                                  &   &                                                    & \cong &                                                                \\
                                  & C &                                                    &       & C                                                             
\end{tikzcd}
\end{center}
In this case, trader $A$ and $B$ are both considered to be wash traders and thus we classify them as a single wash-trading group that either makes a profit or takes a loss through their transaction with legitimate trader $C$. To calculate the profit, we gather the revenue and expense of each trader and wash group.

Filtering through every transaction, we obtained a complete list of disjoint wash trading groups. We collapsed \LooksRareWashtraderCount wash traders on LooksRare into \LooksRareWashgroupCount disjoint wash trading groups. As for Blur, \BlurWashtraderCount wash traders were categorized into \BlurWashgroupCount wash groups. Similarly, on OpenSea, \OpenSeaWashtraderCount wash traders were grouped into \OpenSeaWashgroupCount disjoint wash trading groups. Excluding the traders who took a loss and those with profits over the 99th percentile to avoid extreme values, Figure~\ref{fig:trader_profit_comparison} compares the wash trading group to legitimate trader profits:

\begin{figure}[H]
\begin{adjustwidth}{-1cm}{-1cm}
\begin{center}
  \begin{subfigure}{.35\textwidth}
    \centering
    \includegraphics[width=1\linewidth]{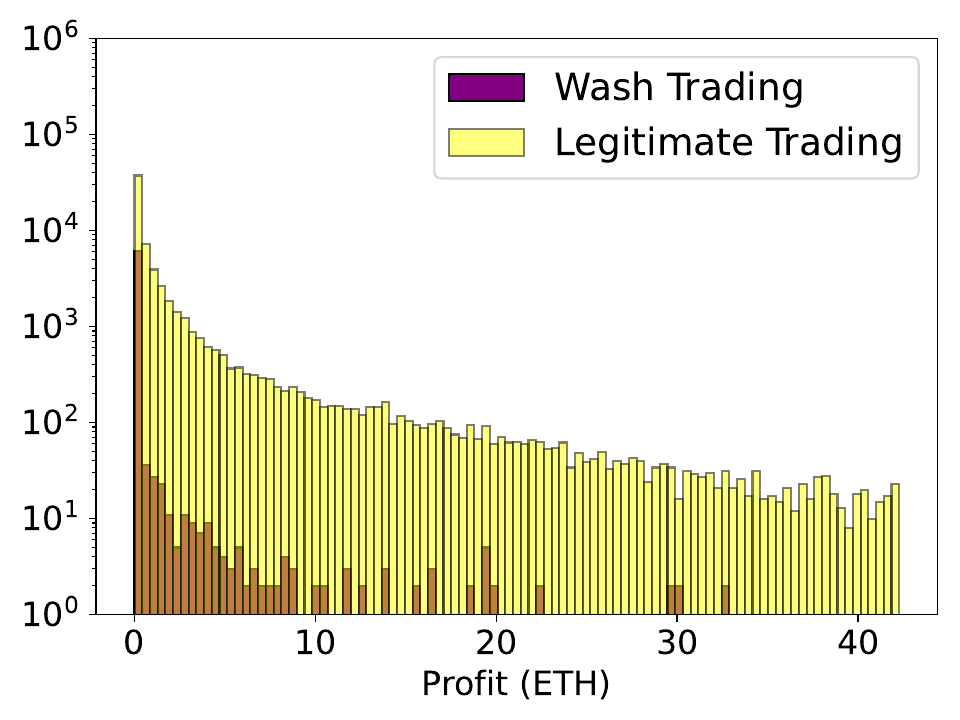}
    \caption{LooksRare}
    \label{fig:LR_profit}
  \end{subfigure}%
  \hspace{-0.1cm}%
  \begin{subfigure}{.35\textwidth}
    \centering
    \includegraphics[width=1\linewidth]{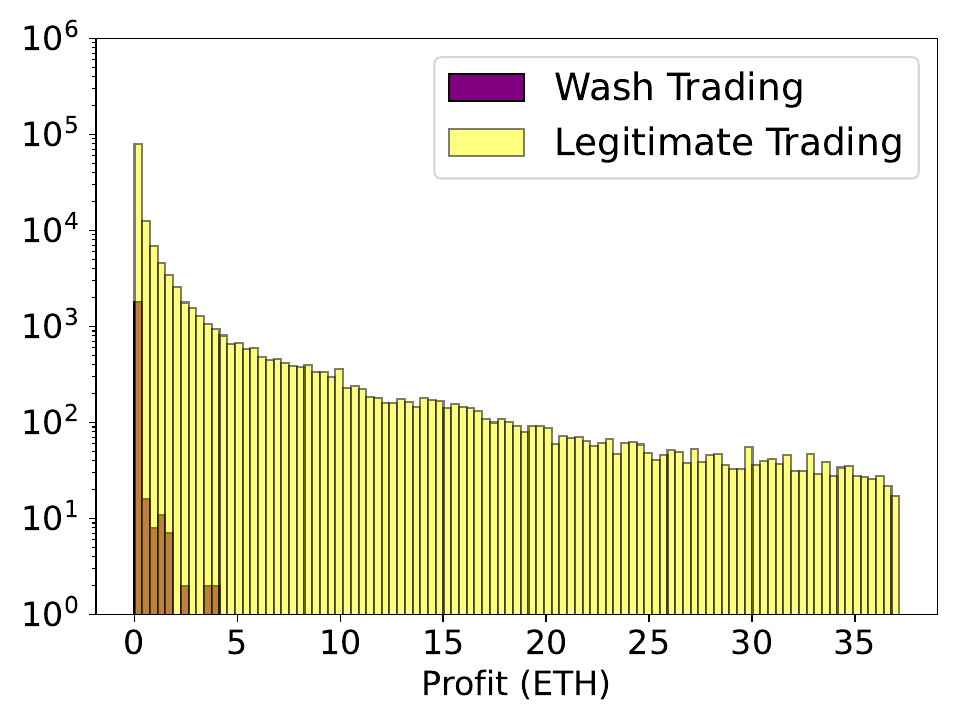}
    \caption{Blur}
    \label{fig:Blur_profit}
  \end{subfigure}%
  \hspace{-0.1cm}%
  \begin{subfigure}{.35\textwidth}
    \centering
    \includegraphics[width=1\linewidth]{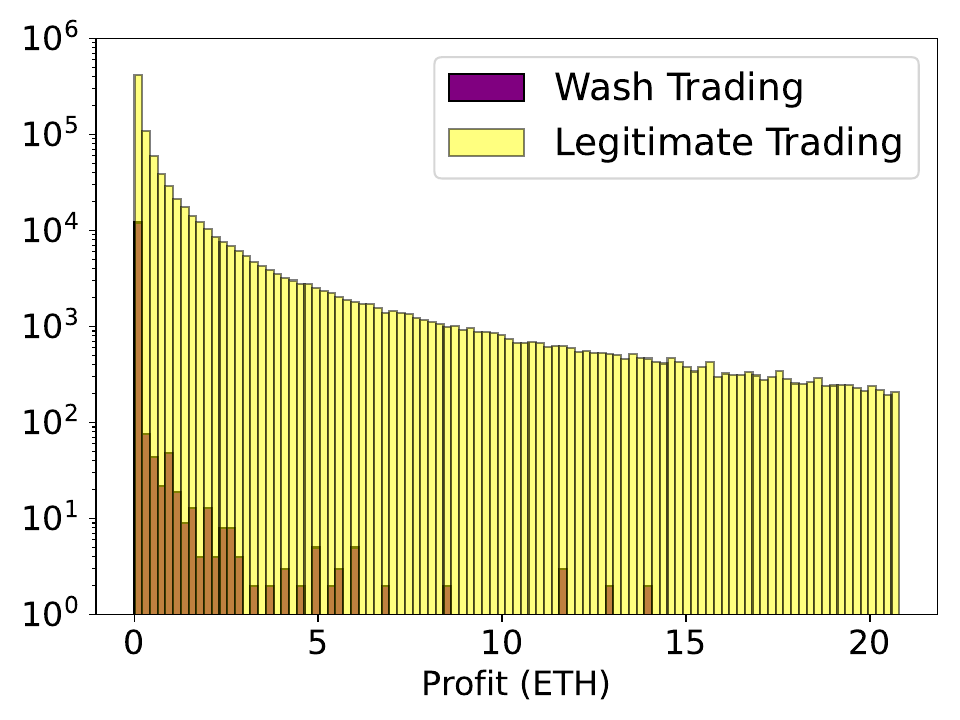}
    \caption{OpenSea}
    \label{fig:OpenSea_profit}
  \end{subfigure}
\end{center}
\end{adjustwidth}
\caption{Distributin of trader profits.}
\label{fig:trader_profit_comparison}
\end{figure}

These histograms visualize the profits earned by traders involved in wash trading versus those engaging in legitimate trading across both platforms. In each histogram, the x-axis represents the profit range in ETH, while the y-axis shows the number of traders or wash trading groups on a logarithmic scale, enabling a clear comparison of the frequency of profit amounts. 

The histograms reveal that, across various profit brackets, wash traders do not consistently secure higher profits when compared to their legitimate counterparts. This visual evidence is corroborated by the following numbers:

On LooksRare, the median profit for wash trading groups is approximately zero ETH, with a maximum profit reaching \LooksRareWprofitMax ETH. In contrast, legitimate traders attain a higher median profit of \LooksRareNWprofitMedian ETH and achieve up to \LooksRareNWprofitMax ETH at the higher end. 

On Blur, wash trading groups achieve a median profit close to zero ETH, with their most profitable outcomes peaking at \BlurWprofitMax ETH. This is lower than the median profit for legitimate traders, which stands at \BlurNWprofitMedian ETH, and legitimate traders also reach much more substantial maximum profits of \BlurNWprofitMax ETH. 

On OpenSea, both wash trading groups and legitimate traders achieve a median profit close to zero ETH. However, legitimate traders can achieve significantly higher maximum profits, reaching up to \OpenSeaNWprofitMax ETH, compared to wash trading groups with a maximum profit of \OpenSeaWprofitMax ETH. These findings suggest that while wash trading may influence market dynamics, it does not necessarily result in greater profitability for those who engage in it.

\subsection*{Summary of Direct Estimation Results}

\begin{enumerate}
    \item[(I1)] Wash trading is identified as a percentage of total trades \LooksRareFilallfourFrac\% on LooksRare, \BlurFilallfourFrac\% on Blur, and \OpenSeaFilallfourFrac\% on OpenSea; yet these trades contribute disproportionately to the total value, accounting for \LooksRareFilallfourVol\%, \BlurFilallfourVol\%, and \OpenSeaFilallfourVol\% of the value in ETH, respectively.
    \item[(I2)] NFT collections implicated in wash trading typically show higher median transaction prices when compared to legitimate trading collections, with a more noticeable discrepancy observed on Blur. This indicates a potential influence of wash trading on inflating NFT collection values.
    \item[(I3)] Wash trading activities do not guarantee higher profits. Data reveals that wash traders do not consistently outperform legitimate traders in terms of profit margins.
\end{enumerate}

\section{Indirect Estimation \#1: Exchange-level Regressions}
\label{sec:regressions}

Among the indirect statistical methods analyzed in \cite{cryptowashtrading}, Trade-Size Roundedness stands out for its ability to quantify aggregate wash trading percentages by combining roundness metrics
with a regression-based approach. 

Leveraging our direct estimation methods from Section~\ref{sec:direct_estimation}, we make two meaningful extensions to their method: In Section~\ref{sec:modreg}, we slightly modify their regression to apply it to the NFT setting where the benchmark for ground truth is given by the direct filters. As both approaches use a 1\% roundedness threshold (discussed below), we refer to them as ``1-Regression'' going forward. In section~\ref{sec:taureg}, we further extend this method via hyper-parameter optimization, and show how this can reduce estimation errors between direct and indirect methods. We refer to this approach as ``$\tau$-Regression'' where the threshold $\tau$ is the outcome of an optimization.

\subsection{1-Regression}\label{sec:modreg}

\cite{cryptowashtrading} categorize a trade price as ``round” if its last nonzero digit exceeds 1\% of the total price.\footnote{For instance, $2.1$ is ``round'' because $.1$ is more than $1\%$ of $2.1$.  On the other hand, $1.135$ is not round because the last nonzero digit, $.005$, is less than $1\%$ of $1.135$.}
The authors use this to quantify the amount of unrounded trades that are typical on a hypothetical exchange with exactly zero wash trades. As a proxy, they assume that highly regulated exchanges likely have very low wash trading activity, and could thus serve in lieu of this ideal benchmark. The amount of unrounded volume obtained from these regulated exchanges is referred to as the Benchmark Unrounded Volume (BUV). Deviations from the BUV indicate wash trading activity, and this is captured via the regression Equation~\ref{eq:creg}.
One important difference in our setting is that since our dataset allows for direct trade-level estimation, we can leverage our results from Section~\ref{sec:direct_estimation} to provide a bespoke BUV for each platform, improving regression accuracy.

In particular, for each complete week of data in our dataset, indexed by $t$, we aggregate the weekly rounded and unrounded traded volumes of legitimate non-wash trades (those not tagged by $\cup f_i$), denoted by ${V^{\operatorname{legit}}_{\operatorname{Rounded}}}_t$ and ${V^{\operatorname{legit}}_{\operatorname{Unrounded}}}_t$. We then perform a log-log linear regression for each exchange:
\begin{align} \label{eq:creg}
    \ln\inparen{{V^{\operatorname{legit}}_{\operatorname{Unrounded}}}_t} = c + \beta \cdot \ln\inparen{{V^{\operatorname{legit}}_{\operatorname{Rounded}}}_t} + \epsilon_t
\end{align}
where $c$ is the constant term, $\beta$ is the regression coefficient, and $\epsilon_t$ is the error term.

Once we obtain the estimated parameters $\hat{c}$ and $\hat{\beta}$, they are used to predict the aforementioned BUV, which is the total predicted legitimate unrounded trade volume $\hat{V}^{\operatorname{legit}}_{\operatorname{Unrounded}}$, using the legitimate rounded non-wash trade volume $V^{\operatorname{legit}}_{\operatorname{Rounded}}$ as input. The deviation of the predicted unrounded volume from the actual unrounded volume (which includes both legit and wash trades) captures wash trading activity:
\begin{align}
    \operatorname{Wash Volume} = \max\inset{V_{\operatorname{Unrounded}} - \hat{V}^{\operatorname{legit}}_{\operatorname{Unrounded}}, 0},
\end{align}
with
\begin{align} \label{eq:pred_unround}
    \hat{V}^{\operatorname{legit}}_{\operatorname{Unrounded}} = \exp\inparen{\hat{c} + \hat{\beta} \cdot \ln\inparen{V^{\operatorname{legit}}_{\operatorname{Rounded}}}},
\end{align}
where $\operatorname{Wash Volume}$ represents the \emph{non-negative} excess of total unrounded trade volume over the predicted legitimate unrounded volume, capturing potential wash trading activity. Then, the estimated wash \% can be computed by dividing by total volume traded.

\subsubsection*{1-Regression Results}
Table~\ref{tab:wash_regression_comparison} compares the estimated percentage of wash trading value obtained using the 1-Regression method with the results from the direct estimation method from Section~\ref{sec:direct_estimation}.  The error column represents the absolute difference between the 1-Regression estimates and the direct wash value estimates.

\begin{table}[H]
\centering
\begin{tabular}{lccccc}
\toprule
{}       &  1-Regression    & 1-Regression      & 1-Regression & Direct $\cup f_i$ &  Abs. \\
Platform &  $R^2$           & p-value           & Wash Value \% &  Wash Value \%   & Error \\
\midrule
     Blur & 0.905 &           2.5e-31 &                     12.64\% &                 58.16\% & $45.52\%$ \\
LooksRare & 0.804 &           5.9e-24 &                     76.39\% &                 95.68\% & $19.29\%$ \\
  OpenSea & 0.909 &           4.9e-35 &                     51.92\% &                 26.39\% & $25.53\%$ \\
  \midrule
  Averages &  & & 46.98\% & 60.08\%  & 30.11\%\\
\bottomrule
\end{tabular}%
\caption{Comparison of 1-Regression and direct wash value estimates across platforms. Avg. $1$-regression aggregate error: $60.08\%-46.98\%=13.09\%$. Avg. $1$-regression absolute error: 30.11\%.}
\label{tab:wash_regression_comparison}
\end{table}

As the table suggests, the errors between the two methods, direct and indirect, can be significant. The 1-regression  underestimates wash trading on Blur (-45.2\%). Less so on LooksRare (-19.3\%). On the other hand, it overestimates wash trading on OpenSea (by 25.5\%). Next, we explore whether these errors can be reduced by leveraging on-chain trade-level data.

\subsection{$\tau$-Regression}\label{sec:taureg}

The 1\% roundedness threshold chosen in \cite{cryptowashtrading} is not the result of an optimization process, and it's possible, even likely, that a different threshold may be optimal. Fortunately, the availability of on-chain data and our direct estimation results allow us to vary the 1\% roundedness threshold used in \cite{cryptowashtrading} and  evaluate its impact on the errors between the two methods.

\begin{figure}[H]
    \centering
    \includegraphics[width=0.6\textwidth]{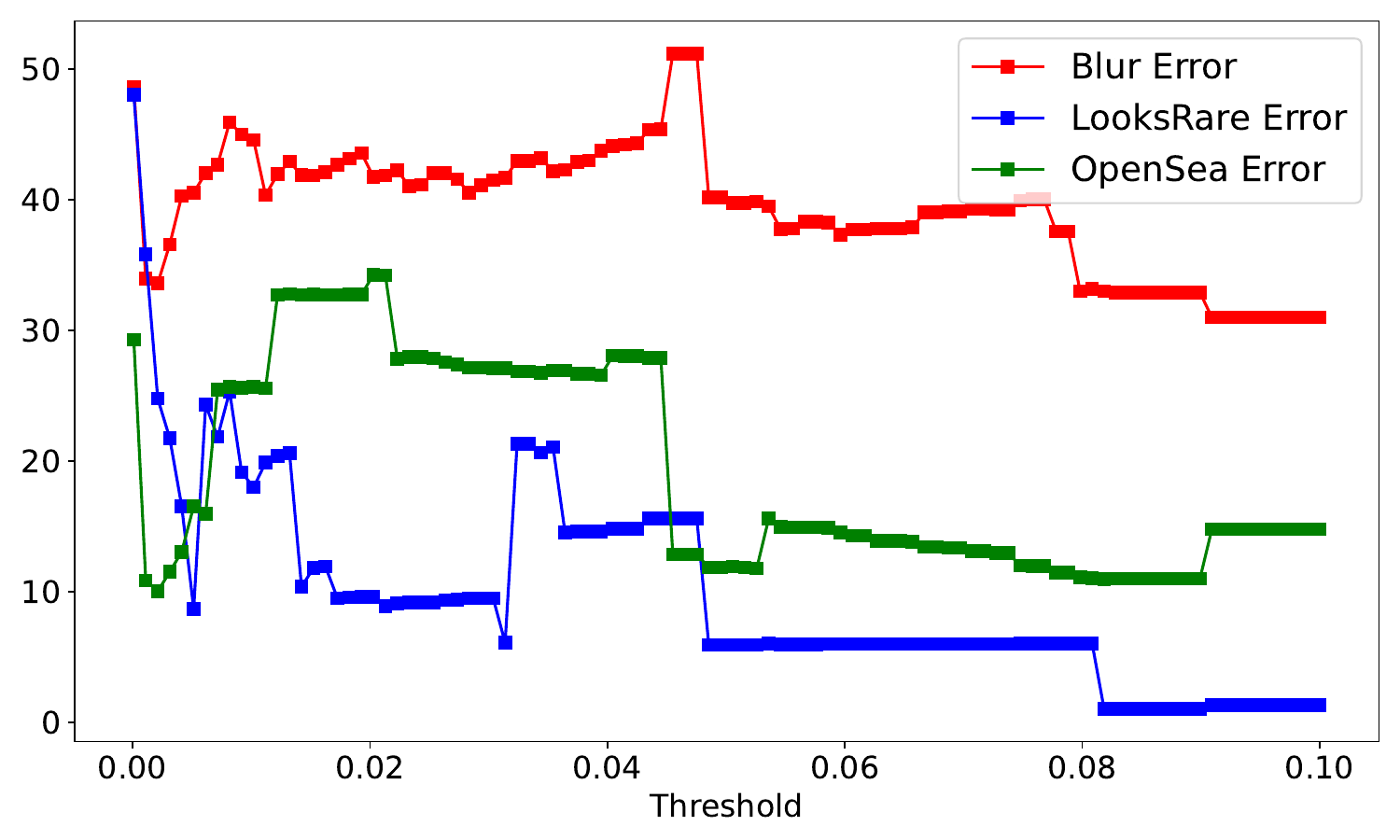}
    \caption{Estimation Error Abs(Direct $-$ Indirect wash \%), versus thresholds $\tau$.}
    \label{fig:regression_varying_threshold}
\end{figure}

We denote the roundedness threshold $\tau$ and to optimize it, we conduct a line-search through 100 evenly spaced values from $0.01\% (0.0001)$ to $10\% (0.1)$, applying the regression method to each one. Figure~\ref{fig:regression_varying_threshold} illustrates the impact of varying thresholds on errors. Each point on the figure represents an independent log-log regression. Thus, we run hundreds of such regressions to populate these figures.

Table~\ref{tab:optimal_thresholds} summarizes the results.
\begin{table}[H]
\centering
\begin{tabular}{lccccc}
\toprule
 {}       & Optimal              & $\tau$-Regression & Direct $\cup f_i$ & $\tau$-Regression  & 1-Regresion \\
 Platform & Threshold ($\tau$) & Wash Value \%     & Wash Value \%     & Abs. Error              & Abs. Error \\
\midrule
     Blur       & 9.09\%  & 27.17\% & 58.16\%& {30.99}\% & {45.52}\% \\
     LooksRare  & 8.18\%  & 94.65\% & 95.68\% & {1.03}\%  & 19.29\% \\
     OpenSea    & 0.21\%  & 36.42\% & 26.39\% & {10.03}\% & 25.53\% \\
\midrule  
Averages &  5.83\% & 52.75\% & 60.08\% & 14.02\%  & 30.11\%\\
\bottomrule
\end{tabular}%
\caption{Comparison of 1-Regression and direct wash value estimates across platforms. Avg. $\tau$-regression aggregate error: $60.08\%-52.75\%=7.33\%$. Avg. $\tau$-regression absolute error: 14.02\%.}
\label{tab:optimal_thresholds}
\end{table}

The optimal threshold is defined as the one yielding the smallest absolute difference, i.e., minimum \emph{error}, between regression-based and direct wash value percentage estimates. Notably, this optimal threshold does not align with the fixed $1\%$ threshold across platforms. The optimal thresholds are 9.09\% for Blur, 8.18\% for LooksRare, and 0.21\% for OpenSea. These optimal thresholds lead to a significant reduction in error between predicted and direct wash value estimates, in some cases by an order of magnitude. For instance, LooksRare's error drops from 19.29\% to just 1.03\%. These observations validate the potential usefulness of the \cite{cryptowashtrading} log-log regression approach for some exchanges, while emphasizing the need for platform-specific fine-tuning. 

\subsection*{Summary of Indirect Estimation Method \#1 - Regressions}

\begin{enumerate}
    \item[(I4)] The standard 1\% threshold regression ``1-regression'' from \cite{cryptowashtrading} shows promise when applied to NFT data, but unoptimized, it can lead to significant estimation errors compared to the direct estimation method.
    \item[(I5)] The optimized $\tau$-regression meaningfully reduces these errors (down to 1.03\% for LooksRare!) but a significant gap still remains for the Blur marketplace.
\end{enumerate}

\section{Indirect Estimation \#2: Machine Learning}\label{sec:ml}

While the $\tau$-Regression approach from Section~\ref{sec:taureg} shows promise in narrowing the gap between direct and indirect estimation methods, notable discrepancies remain---particularly on the Blur platform, where a 7.33\% residual aggregate error (and 30.11\% absolute average error) persists even after fine-tuning the threshold. To address these issues, we propose herein a comprehensive artificial intelligence / machine learning framework. We optimize three models: two tree-based algorithms (random forests and XGBoost) and a deep learning neural network based on the TabNet  architecture \citep{arik2021tabnet}. 

As the first two models are more commonly discussed in the literature, we focus our attention on the latter here.\footnote{complete hyperparameter and training settings are available in the github accompaniment to the paper. A summary is provided in Appendix~\ref{sec:hp}.}
The attention mechanism in TabNet is conceptually similar to the multi-head self-attention used in large language models; while LLMs capture inter-token dependencies, here the mechanism dynamically weights features on an instance-by-instance basis. This capability allows the model to focus on specific combinations of trade volume, timing, and account activity that are indicative of wash trading, even if individual features may not appear anomalous.

Designed with tabular data in mind, the network employs a sequential process in which each decision step selectively attends to a subset of features. An attentive transformer first generates a sparse mask that highlights the most relevant inputs; these selected features are then processed by a feature transformer module—comprising fully connected layers with non-linear activations—to capture complex interactions and conditional dependencies among trading metrics. This approach  preserves some degree of interpretability by revealing which features most influenced each prediction.

The primary parameters of this network  are $n_{steps}$ (the number of sequential decision steps), $n_a$ (the dimension of the attention layer, influencing feature interaction complexity), and $n_d$ (the dimension of the decision layer, shaping the output features at each step). Given the size of our dataset and feature list (see Section~\ref{sec:mlfeat}), we aim for a moderate number of parameters to balance expressiveness and avoid overfitting. We  explore $n_a$ and $n_d$ within the range [8, 32], and $n_{steps}$ in [3, 7]. This approach is results in a model with tens of thousands, or at most, low hundreds of thousands of weights. The relaxation parameter ($\gamma$) and sparsity regularization ($\lambda_{sparse}$) are also tuned. The hyperparameter search employs a combination of grid and random search.

Although fine-tuning is essential, a pivotal aspect is the initial feature engineering. To this end, given the promising results from Section~\ref{sec:regressions}, we start by adapting the $\tau$-regression method to operate at the individual trade level.\footnote{To achieve this, we first estimate the $\tau$-regression parameters using the training data, then for each trade, compute its NFT collection’s cumulative rounded and unrounded value up to that point in time, and apply Equation~\ref{eq:pred_unround} from Section~\ref{sec:regressions} to derive the cumulative wash value percentage.} This adapted $\tau$-regression is combined with 15 additional features that capture key characteristics of wash trading behavior.

\subsection{Input Features}\label{sec:mlfeat}
Each sales record in our dataset includes detailed information such as buyer and seller identities, trade value, the NFT collection and token Id, and the timestamp of the transaction. These rich attributes enabled the construction of the following features for each trade:
\begin{itemize}
    \item \textbf{Price}: Value of the trade in its native token (e.g., 0.8 ETH);
    \texttt{price}.
    \item \textbf{Round Level}: Round level of the trade value \\ \texttt{round\_level}.
    \item \textbf{Collection Cumulative Wash Percentage ($\tau$-regression)}: Fraction of estimated wash value relative to the total trading value of the NFT collection, computed up to the given trade timestamp, estimated using \cite{cryptowashtrading}-style log-log regression with optimal threshold from Section~\ref{sec:regressions}; \\ \texttt{cumu\_wash\_percent\_opt}.
    \item \textbf{Trader Trade Count}: Total number of trades by the trader (buyer or seller) in the preceding 24 hours and preceding 7 days\footnote{Each interval (24 hours and 7 days) is calculated separately for both buyers and sellers, resulting in 4 distinct features.}; \\ \texttt{buyer\_24h\_trade\_count}, \texttt{buyer\_7d\_trade\_count}, \texttt{seller\_24h\_trade\_count}, \\\texttt{seller\_7d\_trade\_count}.
    \item \textbf{Trader NFT Trade Count}: Total number of trades by the trader involving the specific NFT collection in the preceding 24 hours and preceding 7 days\footnote{Similar to \textbf{Trader Trade Count}, this generates 4 distinct features due to the inclusion of both buyers and sellers over two time intervals.}; \\ \texttt{buyer\_24h\_nfttrade\_count}, \texttt{buyer\_7d\_nfttrade\_count}, \texttt{seller\_24h\_nfttrade\_count}, \\\texttt{seller\_7d\_nfttrade\_count}.
    \item \textbf{NFT Trade Count}: Total number of times the NFT collection was traded (by any trader) up to this point in time; 
    \texttt{buyer\_nft\_all\_trade\_count}, \texttt{seller\_nft\_all\_trade\_count}.
    \item \textbf{Price Deviation}: Difference between the current trade price and the last recorded trade price for this specific NFT token;
    \texttt{price\_deviation}.
    \item \textbf{Time Since Last Trade}: Elapsed time since the last trade of this specific NFT token;\\
    \texttt{time\_since\_last\_trade}.
    \item \textbf{Time of Day}: Hour of the day when the trade occurred;
    \texttt{hours}.
\end{itemize}
To ensure data integrity, we excluded samples with missing feature values. After cleaning, Blur retains 3,450,059 samples (99.66\% of total), LooksRare retains 401,105 samples (100.00\%), and OpenSea retains 21,309,625 samples (99.69\%). The cleaned data was split into train-test sets with stratification to preserve the distribution of the target label.

Note, we must deliberately exclude all direct filters from Section~\ref{sec:direct_estimation} from our feature set for two reasons. First, and most obviously, these filters are meant to detect direct evidence of wash trading and thus serve as ground truth labels---their inclusion in the feature set would compromise model integrity.\footnote{If an alternative ground truth—such as court orders or leaked private data (e.g., \cite{BitcoinWashTrading})—were available, these filters could be incorporated in the feature set to further improve performance.} Second, we aim to design this ML-based framework in a generalizable way, for applicability beyond NFT markets; the feature set is thus constructed to handle both fungible and non-fungible assets with minimal modifications. The performance of the AI-based estimation approach on NFTs is discussed in \ref{sec:mlresults}, while its performance on fungible tokens is discussed in Section~\ref{sec:aloosh}.

\subsection{ML Results}\label{sec:mlresults} 

We evaluate model predictive performance out-of-sample at two levels: the individual trade level and the aggregate wash trading volume level. Trade-level performance is measured using ROC-AUC scores, while aggregate performance assesses how well each ML algorithm predicts the ``true'' percentage of total wash trading volume (from Section~\ref{sec:direct_estimation}). The aggregate error is quantified as:

\texttt{aggr\_pred\_error} \ =  \ \texttt{true\_wash\_trading\_volume\_\%} \ - \ \texttt{predicted\_wash\_trading\_volume\_\%}

At the trade level, ML models predict the probability of each trade being a wash trade. A binary classification (\texttt{wash} or \texttt{non-wash}) is assigned based on a probability threshold (set at 50\% by default). To evaluate trade-level performance, we use out-of-sample ROC-AUC scores because they are threshold-independent, allowing for an objective comparison across models without requiring a fixed probability-to-label conversion threshold.

In contrast, evaluating \texttt{aggr\_pred\_error}  requires selecting a conversion threshold, and this is typically optimized against a specific objective. The choice of objective is inherently subjective and application-dependent, with possible criteria including minimizing wash number or wash value percentage errors, maximizing F1 or MCC scores, etc. Some contexts may require more elaborate objectives, such as cost-benefit matrices. For our analysis, to keep things simple, we optimize for the threshold that minimizes  \texttt{aggr\_pred\_error}. Table~\ref{tab:pred_true_error} summarizes the out-of-sample performance of each ML model in terms of AUC scores and absolute \texttt{aggr\_pred\_error} (denoted  ``Error'' in the table).

\begin{table}[H]
\centering
\begin{tabular}{@{}lccc|ccc@{}}
\toprule
         & XGB & RF  & DL & {XGB} & {RF} & {DL}\\
Platform & ROC-AUC     & ROC-AUC  & ROC-AUC& {Abs. Error} & {Abs. Error} & {Abs. Error}\\
\midrule
Blur       & 0.908 & 0.721 & 0.846 & 0.02\% & 1.01\% & 0.18\%\\
LooksRare  & 0.847 & 0.746 & 0.790 & 0.01\% & 0.02\% & 0.09\%\\
OpenSea    & 0.778 & 0.647 & 0.674 & 0.12\% & 1.28\% & 2.15\%\\
\bottomrule
\end{tabular}
\caption{Out-of-sample machine learning results. Average ML error across all exchanges and models is $0.54\%$. Legend: XGB=xgboost, RF=random forest, DL=Deep Learning.}
\label{tab:pred_true_error}
\end{table}

The results in Table~\ref{tab:pred_true_error} demonstrate that all three models achieve exceptional performance in reducing aggregate out-of-sample estimation errors. These range from just $0.01\%$ (XGB) to $2.15\%$ (RF). Thus the ML approach substantially outperforms previous regression-based methods. Yet, these regressions remain highly relevant! As a reminder, they now form a critical part of the input feature-set, helping achieve these remarkably low error rates, well beyond their standalone capabilities.

For clarity, we summarize how the error rates drop significantly from $1$-Regression to $\tau$-Regression to the average rates for the three machine learning models, in Table~\ref{tab:error_recap}.

\begin{table}[H]
\centering
\begin{tabular}{@{}lccc@{}}
\toprule
         & 1-Regression & $\tau$-Regression & avg. ML \\
Platform & Abs. Error & Abs. Error & Abs. Error \\
\midrule
Blur       & 45.6\% & 31.0\% & 0.4\% \\
LooksRare  & 19.3\% & 1.0\% & 0.04\% \\
OpenSea    & 25.5\% & 10.0\% & 1.2\% \\
\bottomrule
\end{tabular}
\caption{Absolute Estimation Errors Across Platforms and Methods.}
\label{tab:error_recap}
\end{table}

Regarding trade-level AUC scores, however, the results reveal greater complexity, reinforcing our earlier observation about dispersion between NFT exchanges (as discussed in the direct estimation section \ref{sec:direct_estimation}). 
While all models demonstrate discriminative ability beyond random chance, XGB exhibits a marked advantage, achieving average AUC scores of $0.84$ across platforms and exceeding $0.9$ for Blur specifically. The deep learning model comes in second, though our experiments indicate the training can be further improved if additional compute is available, e.g., by leveraging higher-end GPU clusters and longer sessions (we trained the model using individual mid-tier NVIDIA L40s, A100, and RTX 4090 GPU cards, limited to overnight runs).

\subsubsection*{Factors Driving Wash Trading}
While the algorithms we use are not as interpretable as other ML techniques, we can nonetheless extract importance scores to tag the features that seem to consistently rank high in terms of predictive power. Table~\ref{tab:iscores} summarizes these, ranked by average importance score across all tests, with the top 4 features being
\texttt{time\_since\_last\_trade}, \texttt{buyer\_7d\_trade\_count}, 
\texttt{buyer\_7d\_nfttrade\_count}, and \texttt{price\_deviation}.

\begin{table}[H]
\centering
\small
    \resizebox{\textwidth}{!}{  
\begin{tabular}{llccccccccc|c}
\toprule
     &  & \multicolumn{3}{c}{Blur} & \multicolumn{3}{c}{LooksRare} & \multicolumn{3}{c}{OpenSea} & \\
    \cmidrule(lr){3-5} \cmidrule(lr){6-8} \cmidrule(lr){9-11}
    Rank & Feature & RF & XGB & DL & RF & XGB & DL & RF & XGB & DL & Average \\
    \midrule
1 & \textbf{time\_since\_last\_trade} & 0.082 & 0.099 & 0.226 & 0.118 & 0.111 & 0.034 & 0.502 & 0.315 & 0.025 & 0.168 \\
2 & \textbf{buyer\_7d\_trade\_count} & 0.253 & 0.343 & 0.217 & 0.146 & 0.130 & 0.256 & 0.007 & 0.024 & 0.051 & 0.159 \\
3 & \textbf{price\_deviation} & 0.003 & 0.020 & 0.000 & 0.044 & 0.064 & 0.027 & 0.138 & 0.178 & 0.401 & 0.097 \\
4 & \textbf{buyer\_7d\_nfttrade\_count} & 0.103 & 0.023 & 0.067 & 0.139 & 0.315 & 0.082 & 0.003 & 0.034 & 0.002 & 0.085 \\
\midrule
5 & buyer\_nft\_all\_trade\_count & 0.148 & 0.127 & 0.075 & 0.110 & 0.029 & 0.081 & 0.002 & 0.012 & 0.066 & 0.072 \\
6 & price & 0.031 & 0.044 & 0.070 & 0.130 & 0.105 & 0.179 & 0.020 & 0.034 & 0.000 & 0.068 \\
7 & seller\_nft\_all\_trade\_count & 0.036 & 0.058 & 0.025 & 0.035 & 0.026 & 0.099 & 0.104 & 0.096 & 0.101 & 0.064 \\
8 & seller\_7d\_trade\_count & 0.091 & 0.098 & 0.106 & 0.020 & 0.025 & 0.022 & 0.041 & 0.071 & 0.004 & 0.053 \\
9 & cumu\_wash\_percent\_opt & 0.004 & 0.018 & 0.027 & 0.011 & 0.022 & 0.019 & 0.027 & 0.032 & 0.239 & 0.044 \\
10 & seller\_7d\_nfttrade\_count & 0.026 & 0.042 & 0.007 & 0.037 & 0.052 & 0.010 & 0.083 & 0.053 & 0.033 & 0.038 \\
11 & buyer\_24h\_trade\_count & 0.120 & 0.022 & 0.015 & 0.086 & 0.033 & 0.003 & 0.001 & 0.020 & 0.005 & 0.034 \\
12 & seller\_24h\_trade\_count & 0.018 & 0.030 & 0.034 & 0.010 & 0.018 & 0.063 & 0.022 & 0.036 & 0.074 & 0.034 \\
13 & buyer\_24h\_nfttrade\_count & 0.076 & 0.013 & 0.037 & 0.092 & 0.016 & 0.024 & 0.003 & 0.022 & 0.000 & 0.031 \\
14 & seller\_24h\_nfttrade\_count & 0.006 & 0.020 & 0.007 & 0.018 & 0.029 & 0.075 & 0.044 & 0.018 & 0.000 & 0.024 \\
15 & hours & 0.001 & 0.028 & 0.081 & 0.000 & 0.007 & 0.000 & 0.000 & 0.034 & 0.000 & 0.017 \\
16 & round\_level & 0.001 & 0.013 & 0.008 & 0.005 & 0.019 & 0.025 & 0.003 & 0.021 & 0.000 & 0.011 \\
    \bottomrule
    \end{tabular}
    }
    \caption{Feature scores by model and exchange, sorted by average score}
    \label{tab:iscores}
    \end{table}

Across all three NFT platforms, our models reveal a unifying theme: wash trading detection hinges on identifying abnormal deviations in trading behavior---whether these manifest as sustained, inflated trading volumes or as sudden shifts in trade recency and price dynamics. In every case, the predictive models highlight anomalies in key features that depart from normal market behavior. Notably, 7-day buyer activity, trade recency (as measured by time since the last trade) and price deviation emerge as critical indicators, albeit with differing absolute values across platforms.

Delving into the specifics, on Blur the buyer-related metrics dominate the signal. The buyer 7-day trade count stands out as the top feature (with RF and XGBoost importance values of 0.253 and 0.343, respectively), accompanied by other buyer metrics such as buyer NFT all trade count. Although features like time since last trade (0.082 and 0.099) and various seller metrics contribute as well, the overall pattern on Blur points to wash trading being executed via sustained, artificially elevated buyer activity.

For LooksRare, a similar buyer-centric pattern is evident, though with a slight shift in emphasis. Here, the buyer 7‑day NFT trade count is particularly prominent (0.139 for RF and 0.315 for XGBoost), and the buyer 7-day trade count also plays a key role (0.146 and 0.131). These features suggest that prolonged abnormal buyer activity is a core mechanism for fabricating market liquidity on this platform. Moderate contributions from time since last trade and price further support this view.

In contrast, on OpenSea the dynamics differ markedly. The models assign the highest importance to time since the last trade (0.502 for RF and 0.316 for XGBoost) and to price deviation (0.138 and 0.178, respectively). Here, buyer metrics such as the buyer 7‑day trade count are nearly negligible, while seller-related features (e.g., seller NFT all trade count and seller 7‑day trade count) assume greater prominence. This pattern indicates that on OpenSea, wash trading is characterized by rapid bursts of trading activity and aggressive price shifts---suggesting that coordinated seller actions, along with abrupt trade recency signals, are key markers of manipulation.

\subsection{Robustness of ML Estimation Beyond NFTs}\label{sec:aloosh}

To gauge robustness to crypto assets outside of NFTs, we deploy the machine learning approach from Section~\ref{sec:ml} to the Bitcoin dataset from \cite{BitcoinWashTrading}. Each transaction is represented by a single row with the following schema:

\begin{itemize} \item \textbf{Trade ID:} A unique identifier for the trade on the platform.
    \item \textbf{Buyer:} The anonymized ID of the entity purchasing Bitcoin.
    \item \textbf{Seller:} The anonymized ID of the entity selling Bitcoin.
    \item \textbf{Time:} The timestamp of the transaction.
    \item \textbf{Bitcoins:} The amount of Bitcoin traded in the transaction.
    \item \textbf{Price (USD):} The price per Bitcoin in USD.
\end{itemize}

In line with their approach, we focus on a subsample of Mt. Gox trades between June 26, 2011, and May 20, 2013, capturing the platform's active trading period. This subsample contains 5,539,244 transactions, totalling 480MB.

We rely on the same 16 features as before with minor tweaks to tailor to Bitcoin data: 
\begin{itemize}
    \item \textbf{Price}: Value of the trade in its native token (e.g., 0.8 BTC).
    \item \textbf{Round Level}: Round level of the trade value.
    \item \textbf{Market Cumulative Wash Percentage}: fraction of estimated wash trades relative to total BTC traded in the market up to that point. 
    \item \textbf{Trader Trade Count}: Total number of trades by the trader (buyer or seller) in the preceding 24 hours and preceding 7 days. (4 features; time frame x buyer/seller)
    \item \textbf{Trader BTC Trade Count}: Total number of BTC traded by the trader in the preceding 24 hours and preceding 7 days. (4 features; time frame x buyer/seller)
    \item \textbf{BTC Trade Count}: Total number of BTC was traded (by any trader). (2 features)
    \item \textbf{Price Deviation}: Difference between the current trade price and the last recorded trade price in USD per BTC.
    \item \textbf{Time Since Last Trade}: Elapsed time since the last trade.
    \item \textbf{Time of Day}: Hour of the day when the trade occurred.
\end{itemize}
As before, to preserve model integrity, we do not include the direct filters used by \cite{BitcoinWashTrading} in the feature list, because the authors utilized them to build their ``ground truth'' wash data.

The results are presented in Table~\ref{tab:pred_true_error2}.

\begin{table}[H]
\centering
\begin{tabular}{@{}lccc|ccc@{}}
\toprule
         & XGB & RF  & DL & {XGB} & {RF} & {DL}\\
Platform & AUC     & AUC  & AUC & {Abs. Error} & {Abs. Error} & {Abs. Error}\\
\midrule
\textbf{Mt. Gox } & 0.971 & 0.862 & 0.942 & 0.03\% & 0.08\% & 0.12\% \\
\bottomrule
\end{tabular}
\caption{Out-of-sample machine learning results on Mt. Gox dataset. Legend: XGB=xgboost, RF=random forest, DL=Deep Learning.}
\label{tab:pred_true_error2}
\end{table}

As before, the results demonstrate very low error rates across all three models, while AUC scores are also higher, reaching a remarkable 0.971 and 0.942  with the XGB and deep learning models, respectively. This suggests that our ML method not only transfers beyond NFTs but performs even more effectively in this context. A likely explanation is that we're applying it to a single asset (Bitcoin) rather than diverse NFT collections all traded on the same exchange. This implies two potential improvements for future work: enhancing our NFT feature set with collection-specific characteristics, or implementing an unsupervised clustering algorithm before supervised learning. We leave this for future work. 

Below, we print out the importance scores for the XGB method, given it has the lowest error (tied with Deep Learning) and highest AUC.

\begin{table}[H]
\centering
\begin{tabular}{lll}
\toprule
{} &                 Feature &     Score \\
\midrule
1  &    buyer\_7d\_trade\_count &  0.1958 \\
2  &   seller\_7d\_trade\_count &  0.1466 \\
3  &      seller\_all\_btc\_sum &  0.1226 \\
4  &       buyer\_all\_btc\_sum &  0.1187 \\
5  &   \textbf{cumu\_wash\_percent\_opt ($\tau$-regression)} &  0.0607 \\
\midrule
8  &  seller\_24h\_trade\_count &  0.0444 \\
6  &   buyer\_24h\_trade\_count &  0.0431 \\
9  &        buyer\_7d\_btc\_sum &  0.04224 \\
7  &       buyer\_24h\_btc\_sum &  0.04100 \\
10 &       seller\_7d\_btc\_sum &  0.03931\\
11 &      seller\_24h\_btc\_sum &  0.03924 \\
12 &                bitcoins &  0.02587 \\
13 &                   hours &  0.02491 \\
14 &             round\_level &  0.02396 \\
15 &   time\_since\_last\_trade &  0.01616 \\
16 &         price\_deviation &  0.01515 \\
\bottomrule
\end{tabular}
\caption{Importance scores for XGB Features on Mt. Gox dataset.}
\end{table}

Interestingly, the optimized \cite{cryptowashtrading} regressions we developed in Section 5.2 appear 5th out of 16 in the list, labeled \texttt{cumu\_wash\_percent\_opt}. This relatively high ranking demonstrates the effectiveness of this specialized method for detecting wash trading patterns in fungible tokens, even when compared to more general trading behavior features. 

\subsection*{Summary of Indirect Estimation Method \#2 - Machine Learning}

\begin{enumerate}
    \item[(I6)] All three predictive algorithms we test substantially reduce aggregate estimation errors (including for the Blur marketplace). XGBoost and Deep Learning perform better than Random Forest in terms of ROC-AUC scores.
    \item[(I7)] The main features that drive wash trading predictability relate  to unusual buyer activity, time since last trade, and price deviations. The $\tau$-regression feature also plays an important role for some ML methods.
    \item[(I8)] The machine learning framework extends beyond NFTs, achieving remarkable performance on the Bitcoin dataset from \cite{BitcoinWashTrading}, ($<$1\% errors and 0.97 AUC).
\end{enumerate}

\section{Indirect Estimation \#3: Distributional Methods} \label{sec:indirect_estimation}

In this section, we turn our attention to the distributional statistical tests from \cite{cryptowashtrading} that assess market manipulation presence in aggregate, rather than predicting specific wash trading quantities.

\subsection{Benford's Law}
\label{sec:benford_comparison}

Benford’s Law predicts that in many natural datasets, smaller leading digits appear more frequently than larger ones (e.g., 1 appears $\sim30\%$ of the time, while 9 appears $\sim5\%$). Significant deviation from this expected distribution may indicate manipulation, such as price fixing or wash trading. 

\begin{figure}[h]
\centering
\begin{subfigure}{.44\textwidth}
  \centering
  \includegraphics[width=\textwidth]{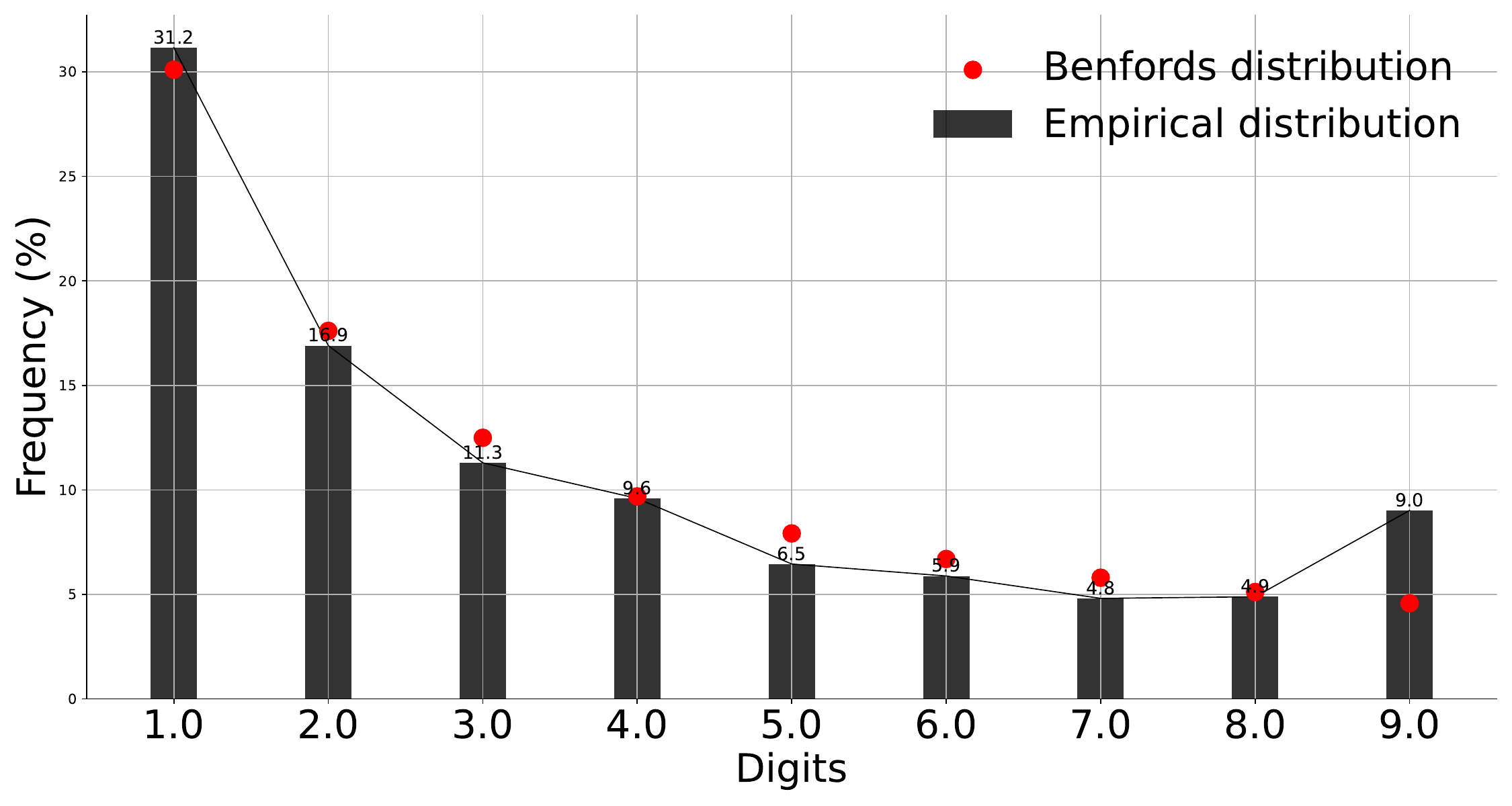}
  \caption{LooksRare Wash Trades. \\Anomaly detected, $p<0.001$, Tstats = 8231.39}
  \label{fig:LR_wash_trades}
\end{subfigure}%
\hspace{0.05\textwidth}
\begin{subfigure}{.44\textwidth}
  \centering
  \includegraphics[width=\textwidth]{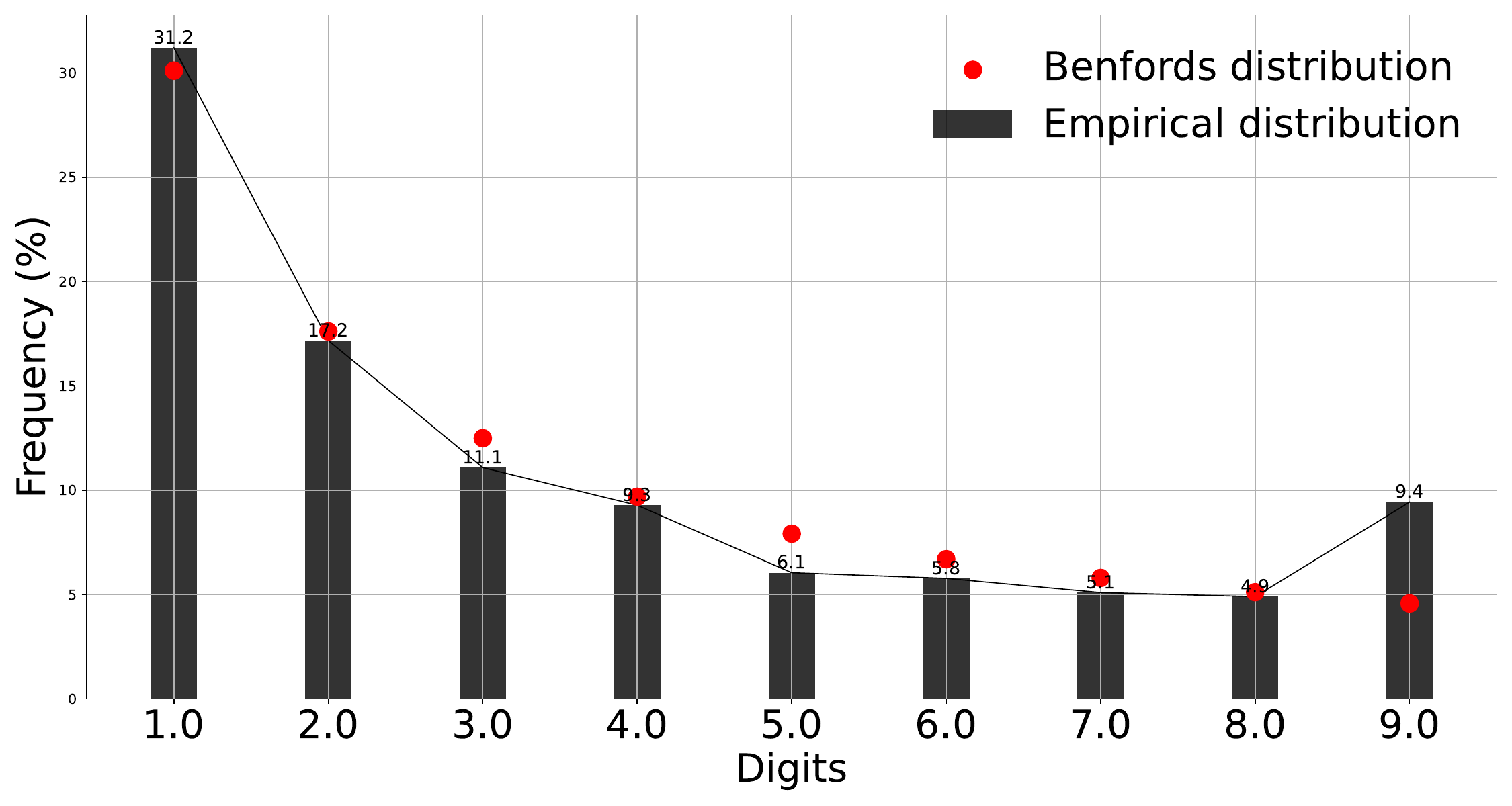}
  \caption{LooksRare Legitimate Trades. \\Anomaly detected, $p<0.001$, Tstats = 14385.7}
  \label{fig:LR_legitimate_trades}
\end{subfigure}
\caption{Distribution of first significant price digits for LooksRare wash and legitimate trades.}
\label{fig:LR_trade_comparison}
\end{figure}

\begin{figure}[h]
\centering
\begin{subfigure}{.44\textwidth}
  \centering
  \includegraphics[width=\textwidth]{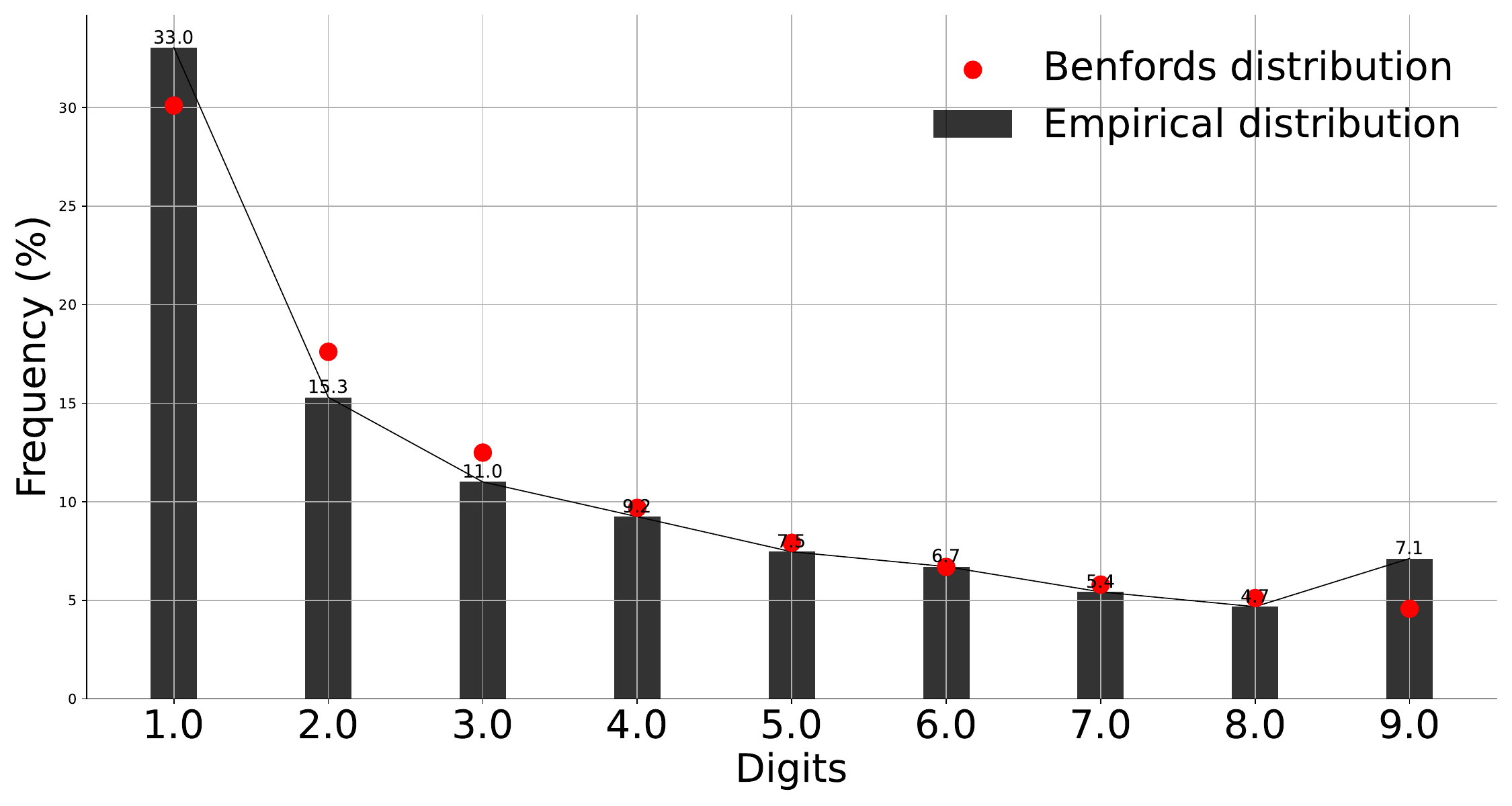}
  \caption{Blur Wash Trades. \\Anomaly detected, $p<0.001$, Tstats = 32744.8}
  \label{fig:Blur_wash_trades}
\end{subfigure}%
\hspace{0.05\textwidth}
\begin{subfigure}{.44\textwidth}
  \centering
  \includegraphics[width=\textwidth]{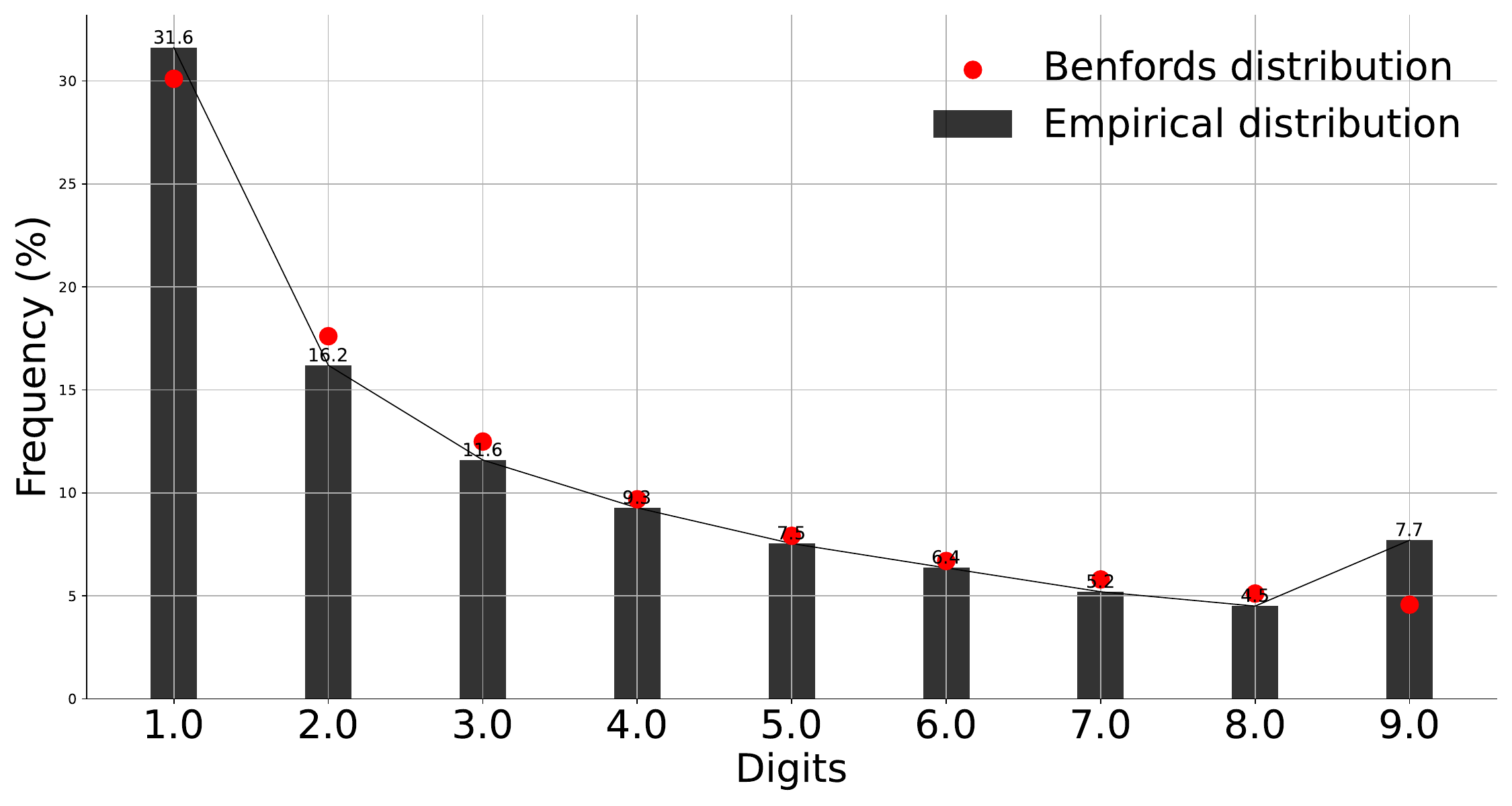}
  \caption{Blur Legitimate Trades. \\Anomaly detected, $p<0.001$, Tstats = 53124.6}
  \label{fig:Blur_legitimate_trades}
\end{subfigure}
\caption{Distribution of first significant price digits for Blur wash and legitimate trades.}
\label{fig:Blur_trade_comparison}
\end{figure}

\begin{figure}[h]
\centering
\begin{subfigure}{.44\textwidth}
  \centering
  \includegraphics[width=\textwidth]{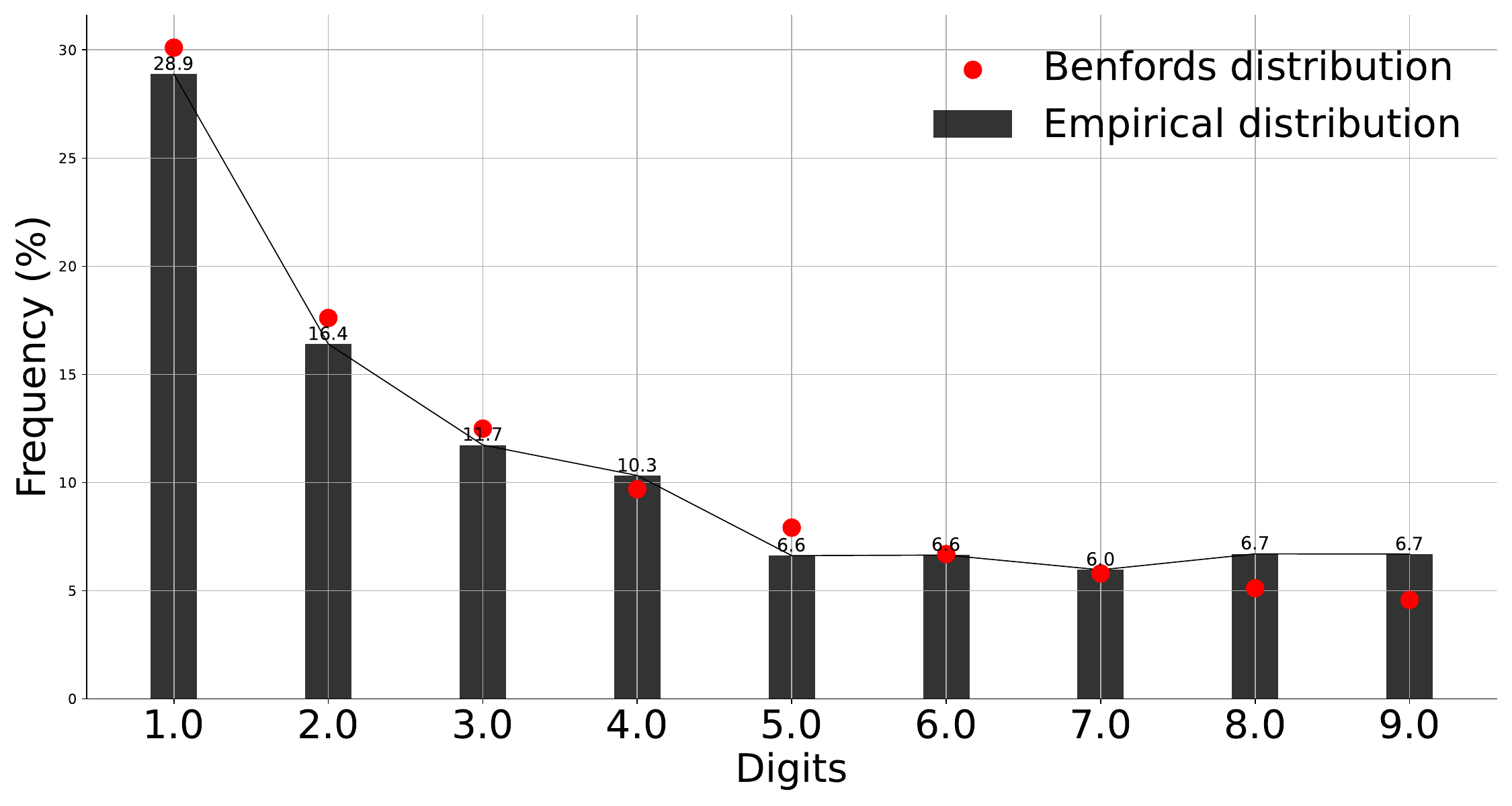}
  \caption{OpenSea Wash Trades. \\Anomaly detected, $p<0.001$, Tstats = 127535}
  \label{fig:OpenSea_wash_trades}
\end{subfigure}%
\hspace{0.05\textwidth}
\begin{subfigure}{.44\textwidth}
  \centering
  \includegraphics[width=\textwidth]{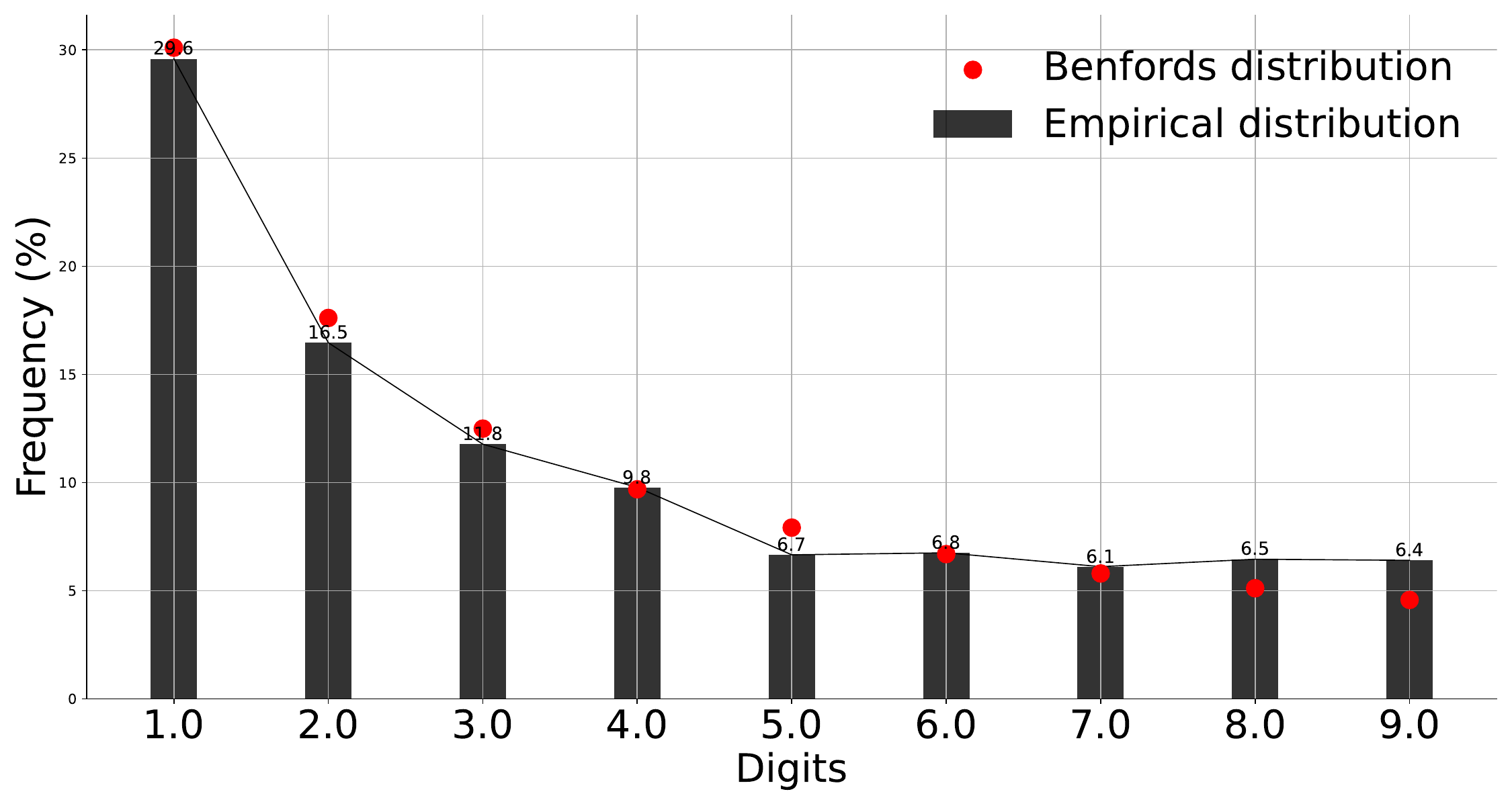}
  \caption{OpenSea Legitimate Trades. \\Anomaly detected, $p<0.001$, Tstats = 211007}
  \label{fig:OpenSea_legitimate_trades}
\end{subfigure}
\caption{Distribution of first significant price digits  for OpenSea wash and legitimate trades.}
\label{fig:OpenSea_trade_comparison}
\end{figure}

For this test, we extract the first nonzero digit from each NFT transaction price (e.g., 1 for 0.125~ETH) and compare it to Benford's expected distribution for wash trades and legitimate trades. We test the null hypothesis that the observed distribution follows Benford's law at a significance level of \(\alpha = 0.05\). Results are shown in Figures~\ref{fig:LR_trade_comparison}, \ref{fig:Blur_trade_comparison}, and \ref{fig:OpenSea_trade_comparison}.

In these figures, the black bars represent the empirical first-digit distribution, while red dots denote the Benford expected values. Deviations indicate non-conformity, quantified by the \(p\)-value (probability of observing such deviation under the null) and the $t$-statistic (magnitude of deviation in standard error units). 

Across LooksRare, Blur, and OpenSea, both wash and legitimate trades yield near zero $p$-values, confirming  statistically significant departures from Benford’s Law. Surprisingly, legitimate trades show higher $t$-statistics than wash trades, indicating even greater divergence. These findings suggest that Benford’s Law may \emph{not} be applicable to NFT data, as both wash and non-wash samples display anomalies. 

This contrasts with \cite{cryptowashtrading}, where regulated exchanges followed Benford’s Law while unregulated Tier-2 exchanges deviated.

\subsection{Trade-Size Clustering}
\label{sec:clustering_comparison}

\begin{figure}[h]
\centering

\begin{subfigure}{.45\textwidth}
  \centering
  \includegraphics[width=1\linewidth]{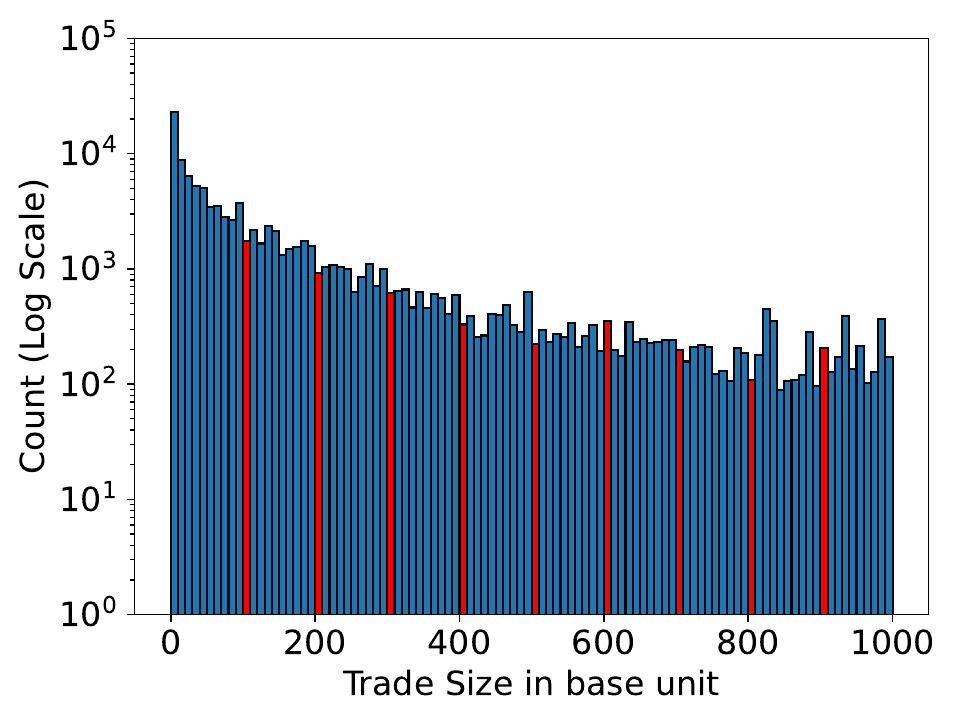}
  \caption{LooksRare Wash Trades}
\end{subfigure}%
\begin{subfigure}{.45\textwidth}
  \centering
  \includegraphics[width=1\linewidth]{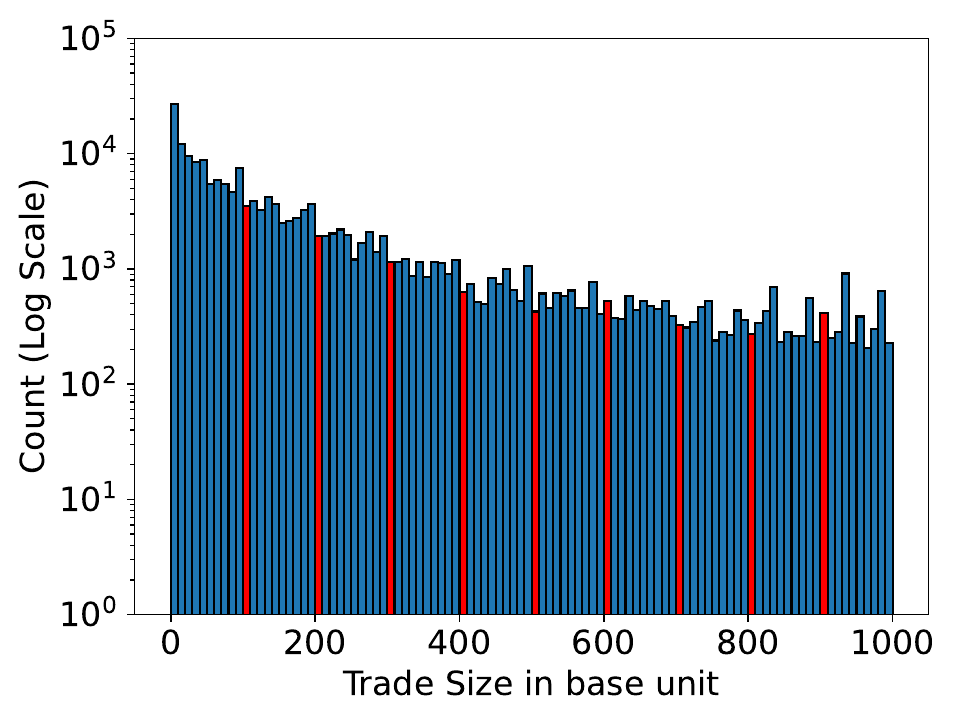}
  \caption{LooksRare Legitimate Trades}
\end{subfigure}

\caption{Trade-size clustering analysis for LooksRare wash and legitimate trades.}
\label{fig:LooksRare_cluster}
\end{figure}

\begin{figure}[h]
\centering

\begin{subfigure}{.45\textwidth}
  \centering
  \includegraphics[width=1\linewidth]{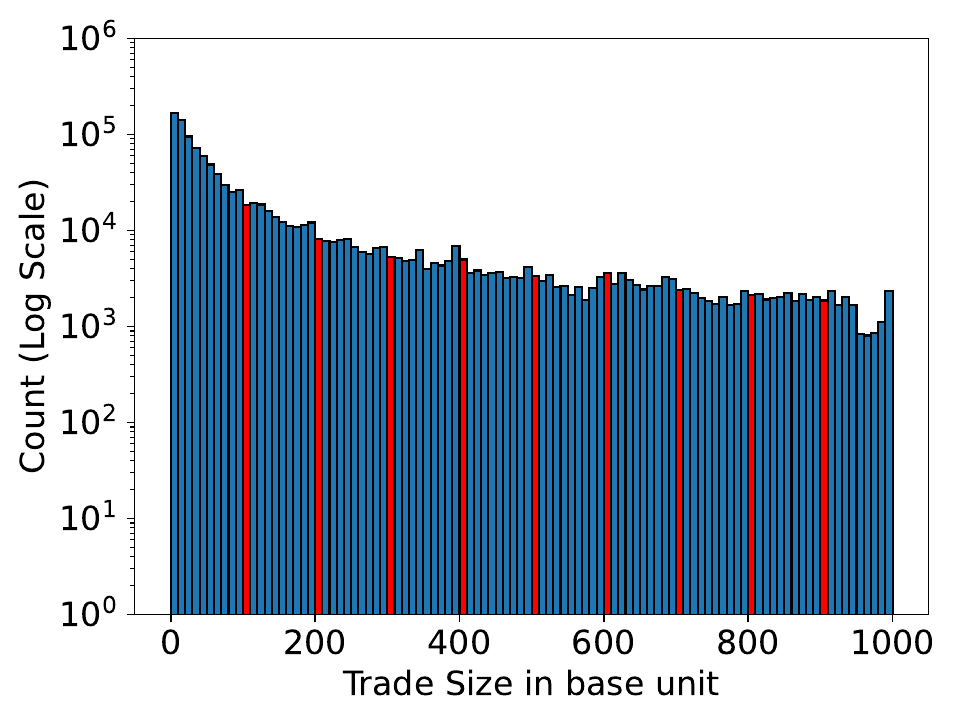}
  \caption{Blur Wash Trades}
\end{subfigure}%
\begin{subfigure}{.45\textwidth}
  \centering
  \includegraphics[width=1\linewidth]{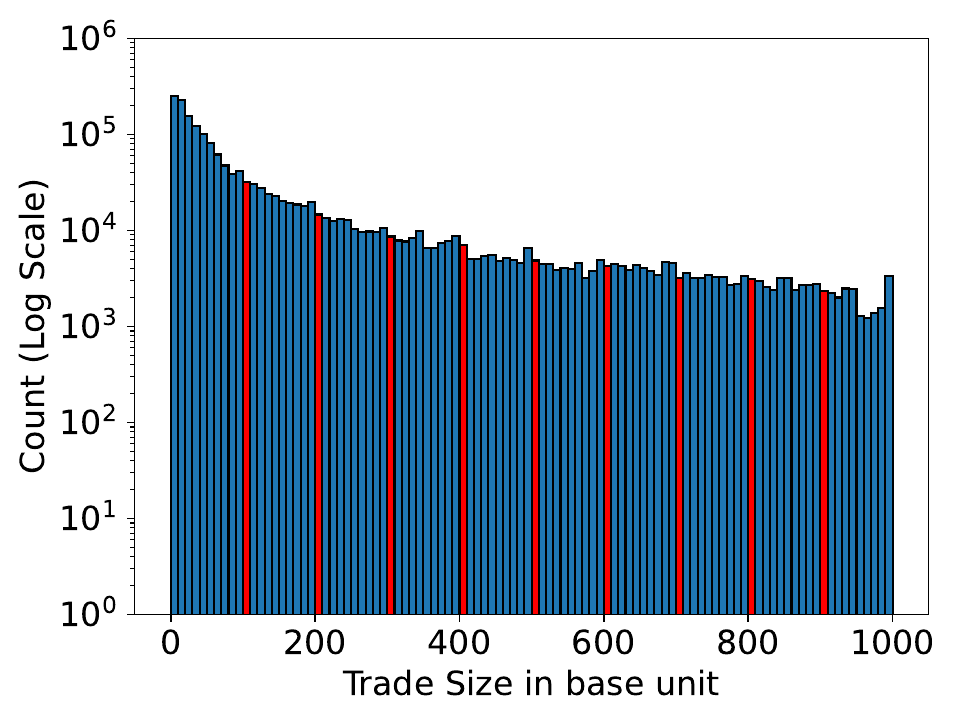}
  \caption{Blur Legitimate Trades}
\end{subfigure}

\caption{Trade-size clustering analysis for Blur wash and legitimate trades.}
\label{fig:Blur_cluster}
\end{figure}

\begin{figure}[h]
\centering
\begin{subfigure}{.45\textwidth}
  \centering
  \includegraphics[width=1\linewidth]{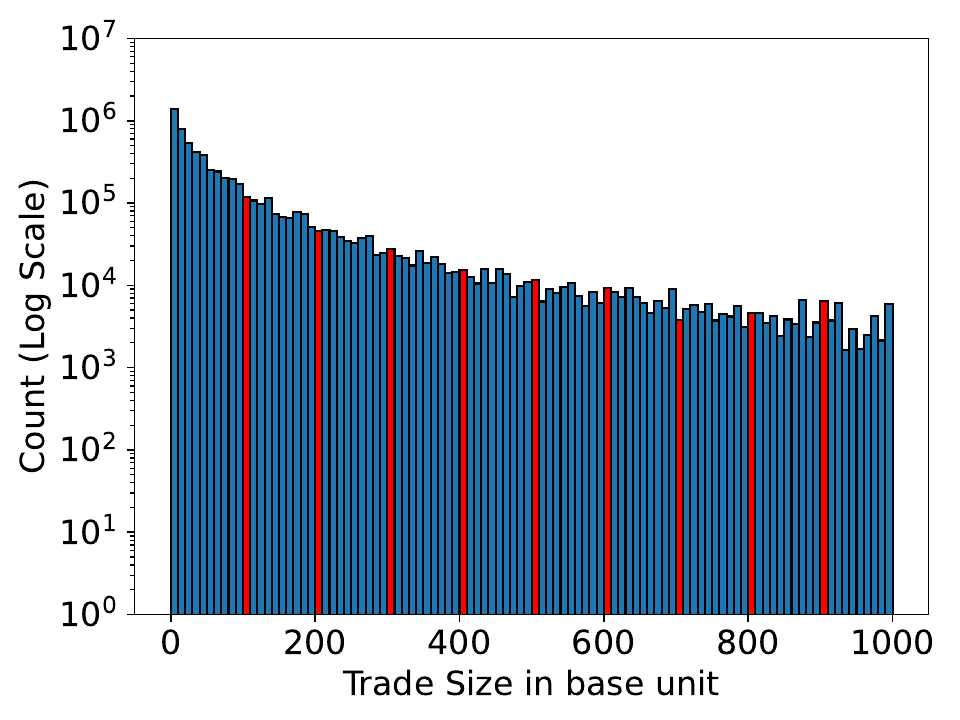}
  \caption{OpenSea Wash Trades}
\end{subfigure}%
\begin{subfigure}{.45\textwidth}
  \centering
  \includegraphics[width=1\linewidth]{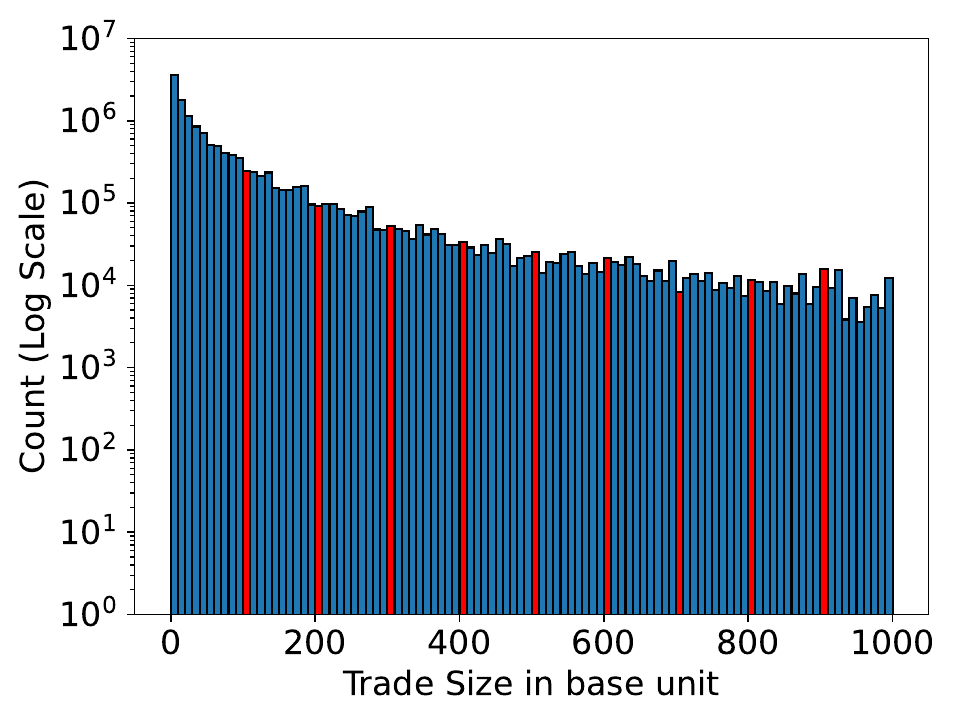}
  \caption{OpenSea Legitimate Trades}
\end{subfigure}

\caption{Trade-size clustering analysis for OpenSea wash and legitimate trades.}
\label{fig:OpenSea_cluster}
\end{figure}

Trade-size clustering stems from the tendency of legitimate traders to prefer round numbers, simplifying decision-making and reducing transaction costs. Following \cite{cryptowashtrading}, we analyze NFT trade prices (in ETH) using a base unit of 0.001 ETH to capture meaningful round sizes. We then assess clustering at multiples of 100 base units to distinguish genuine trades from algorithm-driven wash trading. Results are shown in Figures~\ref{fig:LooksRare_cluster}, \ref{fig:Blur_cluster}, and \ref{fig:OpenSea_cluster}.

These histograms show trade-size distributions on a logarithmic scale. Blue bars represent trade counts per size interval, while red bars mark multiples of 100, highlighting potential clustering. Regular peaks at these intervals could indicate manual or algorithmic trading tendencies, with differences between wash and legitimate trades potentially revealing manipulation.

Across LooksRare, Blur, and OpenSea, we apply a uniform 1000 base-unit cutoff and bin trades in increments of 10. No strong clustering effect is observed in the figures, for either wash or legitimate trades on any platform.

To quantify trade-size clustering more formally, we apply the Student's t-test: 
\[
    t = \frac{\bar{x} - \mu_0}{s/\sqrt{n}},
\]
where \(\bar{x}\) is the difference between the mean frequency of rounded and unrounded trade sizes, \(s\) is the sample standard deviation, and \(n\) is the sample size.

Following \cite{cryptowashtrading}, we compute trade frequencies using two sets of observation windows: 100-unit windows: $\inbrak{100X - 50, 100X +50}$ and 500-unit windows: $\inbrak{500X - 100, 500X + 100}$. A trade is ``rounded” if its size is an exact multiple of 100 (or 500, depending on the window); otherwise, it is ``unrounded.'' The null hypothesis states no difference between the frequencies of rounded and unrounded trades. Table~\ref{tab:tstats_update} presents the \(t\)-statistics for each sub-sample.

\begin{table}[H]
\centering
\begin{tabular}{@{}lcc@{}}
\toprule
\textbf{}                  & \textbf{Windows in 100s} & \textbf{Windows in 500s} \\ \midrule
\textbf{LooksRare wash}    & -16.46                   & -24.00                   \\ 
\textbf{LooksRare legitimate} & -8.53                & -11.06                   \\ 
\textbf{Blur wash}         & -143.56                  & -27.51                   \\ 
\textbf{Blur legitimate}   & -351.77                  & -16.16                   \\ 
\textbf{OpenSea wash}      & -246.86                  & -114.11                  \\ 
\textbf{OpenSea legitimate} & -4.04                   & -1.56                    \\ \bottomrule
\end{tabular}
\caption{$t$-statistics for trade size clustering in 100 and 500 unit windows.}
\label{tab:tstats_update}
\end{table}

All $t$-statistics are negative, indicating that rounded trade sizes occur less frequently than unrounded ones. The absolute value of the $t$-statistic reflects the strength of this effect—larger values indicate stronger avoidance of round numbers.

On LooksRare, both wash and legitimate trades show moderate negative \(t\)-values (wash: \(-16.46\) and \(-24.00\); legitimate: \(-8.53\) and \(-11.06\)), suggesting a limited clustering effect. On Blur, the \(t\)-statistics indicate a strong avoidance of round numbers, especially in the 100-unit window (wash: \(-143.56\); legitimate: \(-351.77\)), while the effect is weaker in the 500-unit window (wash: \(-27.51\); legitimate: \(-16.16\)). On OpenSea, wash trades exhibit a strong avoidance (\(-246.86\) and \(-114.11\)), whereas legitimate trades show only a slight difference (\(-4.04\) and \(-1.56\)).

This method effectively differentiates NFT wash and non-wash trades, though in a subtle way. Unlike \cite{cryptowashtrading}, where regulated exchanges exhibited clear clustering around round numbers, our results indicate a general avoidance of round numbers across all NFT trades. However, legitimate trades still display weaker avoidance than wash trades—except on Blur (100-unit window), where the effect is slightly reversed.

\subsection{Tail Distribution}
\label{sec:tail_comparison}

Financial and cryptocurrency markets often exhibit fat-tailed distributions, well-approximated by power laws:
\[
P(X > x) \sim x^{-\alpha},
\]
where \(\alpha\) is the power-law exponent, reflecting the probability of extreme trade sizes. Large $\alpha$ (steeper slope) indicates thinner tails, while small $\alpha$ (shallower slope) suggests fatter tails.

For NFTs, we analyze the top 10\% of trade sizes (trades above the 90th percentile) and estimate $\alpha$ using Ordinary Least Squares (OLS) and Maximum Likelihood Estimation (MLE) (Hill estimator). In unmanipulated markets, $\alpha$ typically falls in the Pareto-Lévy regime, $\alpha\in \inparen{1,2}$, with deviations suggesting anomalies.

\begin{figure}[h]
\centering
\begin{subfigure}{.45\textwidth}
  \centering
  \includegraphics[width=\linewidth]{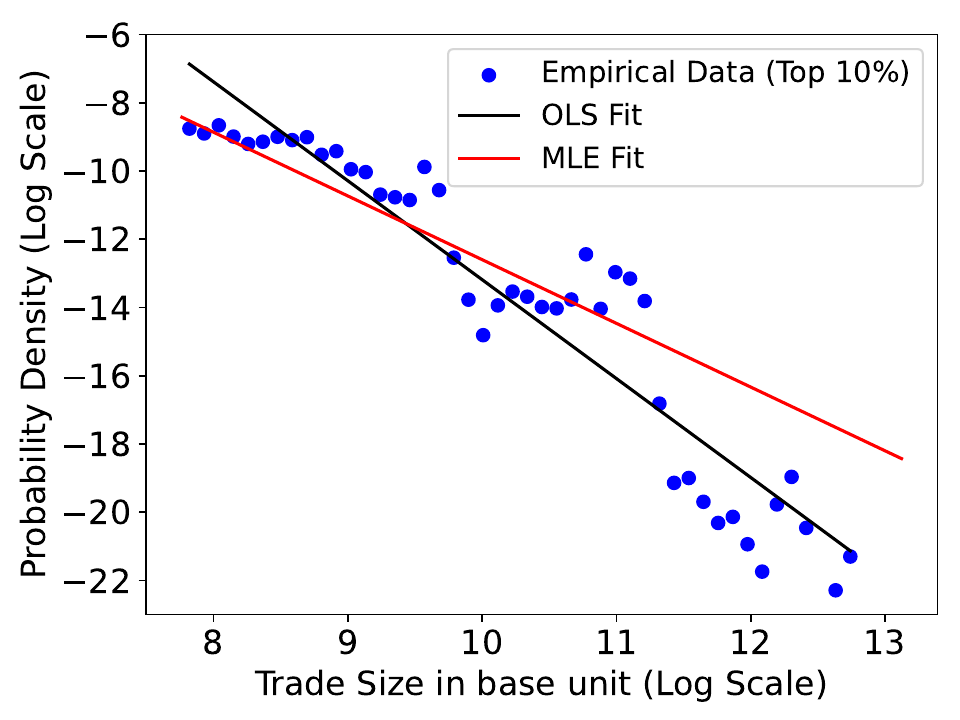}
  \caption{Blur Wash Trades}
\end{subfigure}%
\begin{subfigure}{.45\textwidth}
  \centering
  \includegraphics[width=\linewidth]{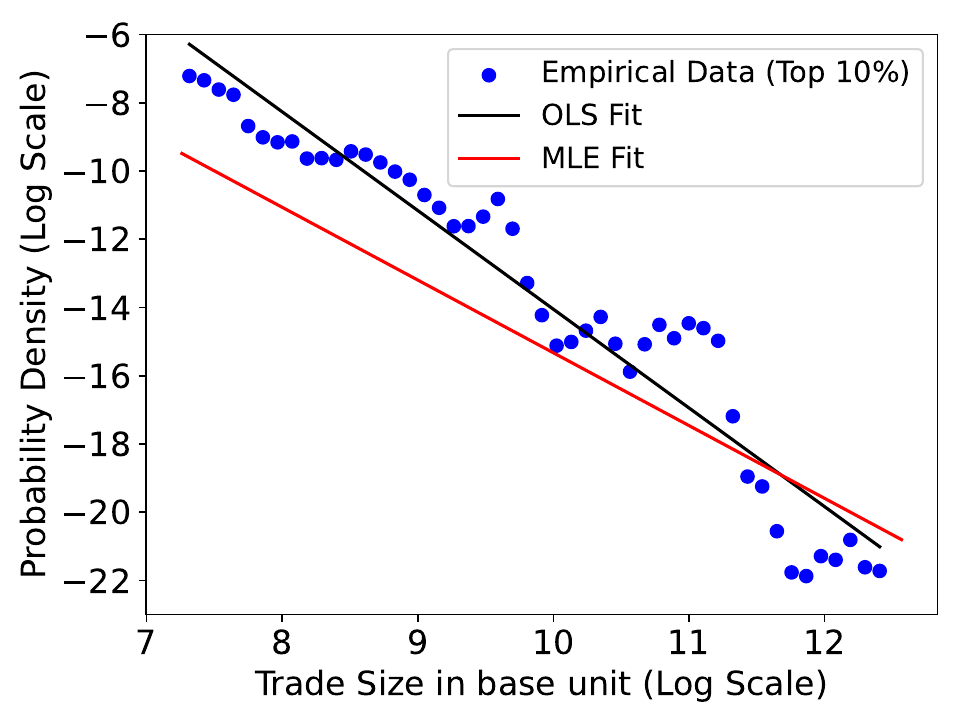}
  \caption{Blur Legitimate Trades}
\end{subfigure}
\caption{Trade-size tail distributions for Blur wash and legitimate trades}
\label{fig:Blur_tail}
\end{figure}

\begin{figure}[h]
\centering
\begin{subfigure}{.45\textwidth}
  \centering
  \includegraphics[width=\linewidth]{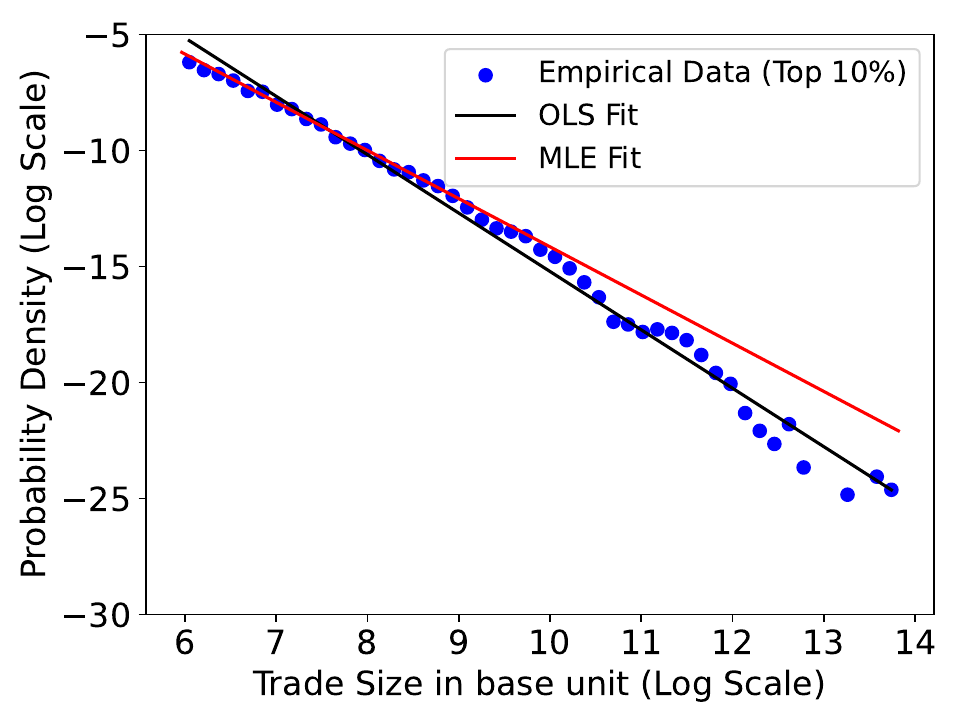}
  \caption{OpenSea Wash Trades}
\end{subfigure}%
\begin{subfigure}{.45\textwidth}
  \centering
  \includegraphics[width=\linewidth]{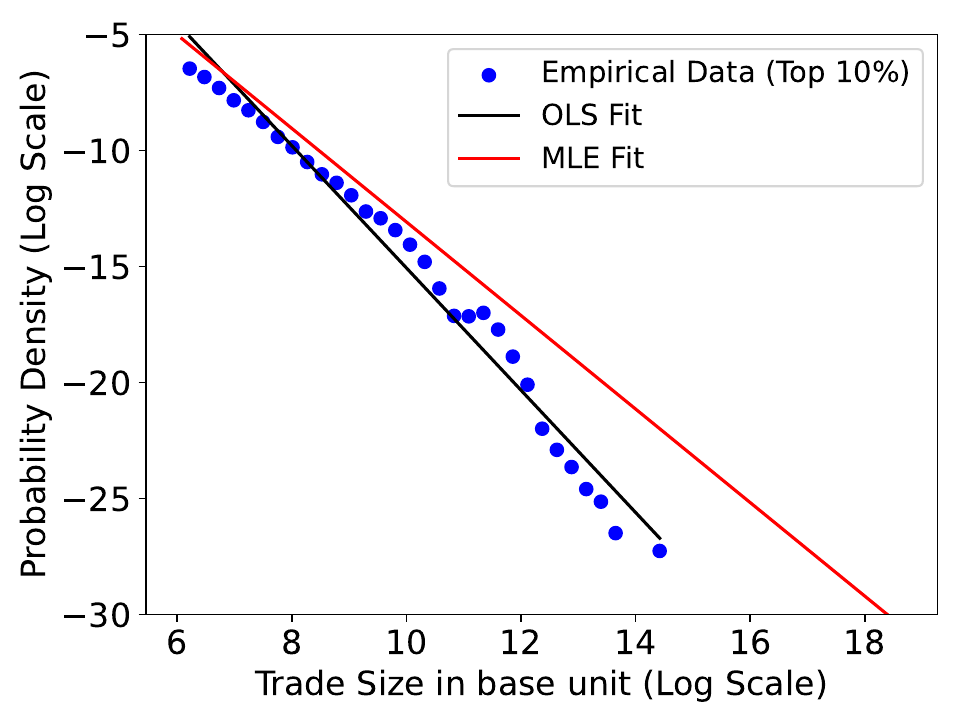}
  \caption{OpenSea Legitimate Trades}
\end{subfigure}
\caption{Trade-size tail distributions for OpenSea wash and legitimate trades}
\label{fig:OpenSea_tail}
\end{figure}

Figures~\ref{fig:LooksRare_tail}, \ref{fig:Blur_tail}, and \ref{fig:OpenSea_tail} compare the tail distributions of wash and legitimate trades across marketplaces. Trade sizes are plotted against probability densities on a log-log scale. Blue dots represent the empirical data, while the red and black lines show the OLS and MLE fits, respectively:

\begin{figure}[h]
\centering
\begin{subfigure}{.45\textwidth}
  \centering
  \includegraphics[width=\linewidth]{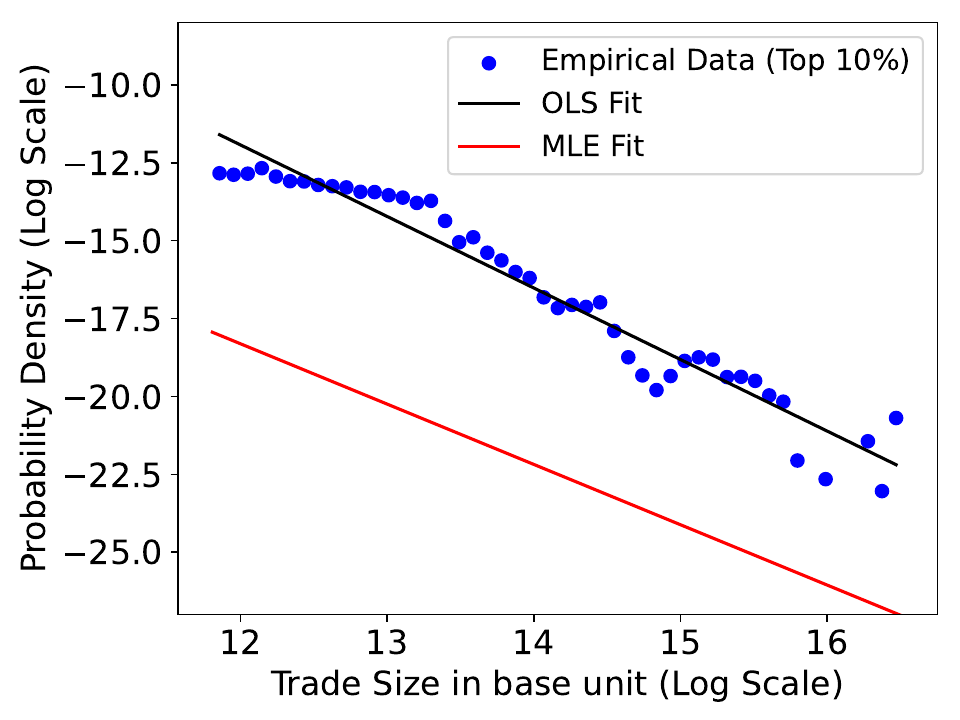}
  \caption{LooksRare Wash Trades}
\end{subfigure}%
\begin{subfigure}{.45\textwidth}
  \centering
  \includegraphics[width=\linewidth]{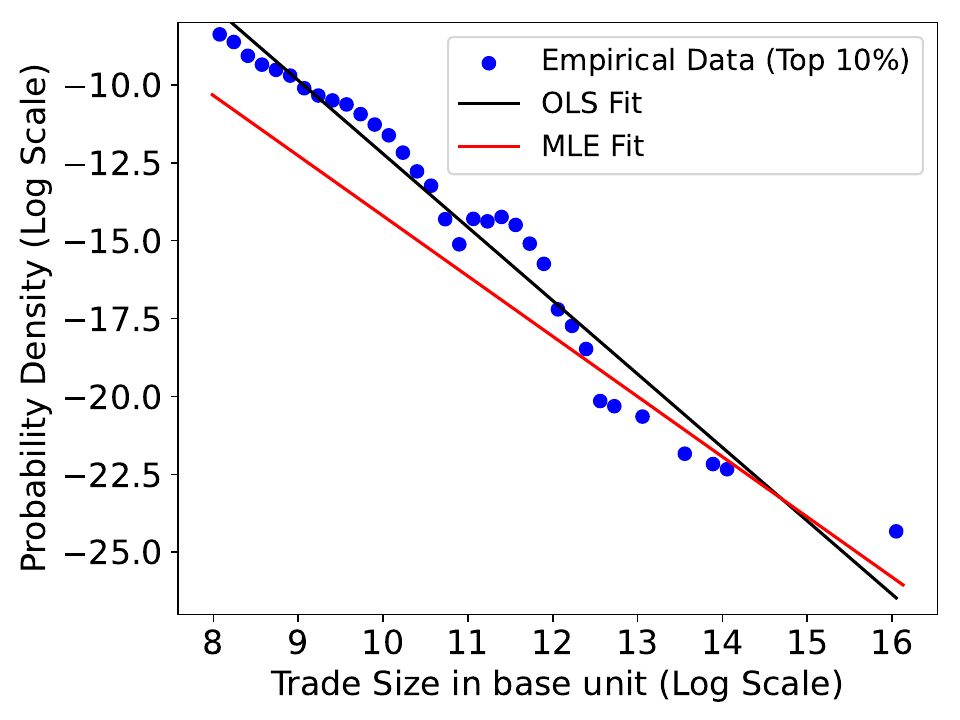}
  \caption{LooksRare Legitimate Trades}
\end{subfigure}
\caption{Trade-size tail distributions for LooksRare wash and legitimate trades}
\label{fig:LooksRare_tail}
\end{figure}

Table~\ref{tab:alpha_exponents} summarizes the estimated exponents. In general, the OLS and MLE lines intersect, except for the LooksRare wash trades where they run parallel—suggesting unique tail behavior, possibly due to outliers or idiosyncrasies in the distribution.

\begin{table}[H]
\centering
\begin{tabular}{@{}lcc@{}}
\toprule
 & \(\hat{\alpha}_{\text{OLS}}\) & \(\hat{\alpha}_{\text{Hill}}\) \\ \midrule
\textbf{LooksRare wash}      & 2.2957 & 1.9357 \\
\textbf{LooksRare legitimate} & 2.3559 & 1.9319 \\
\textbf{Blur wash}           & 2.8993 & 1.8675 \\
\textbf{Blur legitimate}     & 2.8925 & 2.1322 \\
\textbf{OpenSea wash}        & 2.5178 & 2.0780 \\
\textbf{OpenSea legitimate}  & 2.6357 & 2.0164 \\ \bottomrule
\end{tabular}
\caption{Estimated power-law exponents (\(\hat{\alpha}\)) from OLS and Hill Estimator (MLE) for trade-size tails}
\label{tab:alpha_exponents}
\end{table}

\cite{cryptowashtrading} found a clear distinction between regulated and unregulated exchanges: regulated platforms adhered to the Pareto–Lévy regime with tail exponents, $\alpha\in(1,2)$, while unregulated exchanges exhibited deviations, $\alpha\not\in (1,2)$. However, our results show that this method does not apply to NFT markets, as none of the platforms, whether wash trade sub-sample or not, exhibit exponents within the expected range. This suggests that power-law tail analysis may not be a reliable diagnostic for manipulation in NFT markets.

\subsection{Summary of Indirect Estimation Method \#3 - Distributional Methods}

As a summary, the distributional methods yield the following key insights:

\begin{enumerate}
    \item[(I9)] \textbf{Benford's Law:} Both wash and legitimate trades deviate from Benford's Law, with legitimate trades showing greater deviations.
    \item[(I10)] \textbf{Trade-Size Clustering:} Legitimate trades exhibit slightly more clustering around round numbers compared to wash trades.
    \item[(I11)] \textbf{Tail Distribution:} NFT trade-size tails differ from those in traditional financial markets.
\end{enumerate}

\section{Discussion} \label{sec:discussion}

In this paper, we introduced a novel framework for detecting and quantifying wash trading in NFT markets by leveraging public on-chain data from three major exchanges. Our analysis reveals that nearly 38\% of trades---and a disproportionately higher 60\% of traded value---likely involve manipulation. These concise yet robust findings directly challenge the limitations of prior studies that relied on indirect statistical proxies or leaked private data, demonstrating instead the power of granular on-chain evidence.

A central contribution of our work lies in its critical evaluation of existing detection methods. We revisited roundedness-based regressions, as popularized in earlier studies, and showed that while such methods have promise, they are inherently error-prone when applied to NFTs. To overcome these challenges, we developed an enhanced approach: an optimized \(\tau\)-regression method integrated within an AI-driven estimation framework. This hybrid methodology markedly reduces estimation errors---down to below 2.15\%---across various platforms and even generalizes to fungible token markets (Bitcoin). Our study bridges the gap between traditional statistical models and modern machine learning techniques, setting a new benchmark for precision in wash trading detection.

Our findings also shed light on the unique dynamics shaping NFT manipulation. Although NFTs share many of the same wash trading incentives found in cryptocurrency markets, key structural differences---such as on-chain transaction fees, zero-fee trading models, royalty earnings, and price opacity---create distinct economic incentives. These factors, coupled with the unprecedented transparency of blockchain technology, have paradoxically contributed to a market environment where manipulative practices can sometimes thrive. As Web3 innovations continue to redefine the financial landscape, the tension between enhanced transparency and the evolution of sophisticated market manipulation becomes ever more pronounced.

Perhaps most importantly, our work carries significant regulatory implications. The high accuracy of our machine learning models---boasting AUC scores exceeding 0.9 on platforms like Blur and Mt. Gox---indicates that regulators could feasibly implement automated, real-time surveillance systems to flag suspicious trading patterns. Yet, the marked variation in wash trading behavior across exchanges cautions against a uniform, one-size-fits-all regulatory approach. Instead, our results advocate for exchange-specific frameworks to ensure effective detection across diverse market settings. 

Moreover, while our enhanced detection systems provide powerful tools for uncovering manipulative behavior, they are intended to complement, not replace, the judicial process. Some trading behaviors---such as asset transfers conducted for privacy, tax optimization, or risk management purposes---may superficially resemble wash trading, leading to potential misclassification. This inherent ambiguity underscores a broader challenge in wash trading enforcement, which often relies on inferring intent from circumstantial evidence. U.S. regulators, for example, define wash trading based on its outcome—specifically, the absence of genuine ownership or risk change—rather than on explicit intent, making judicial deliberation essential. By following outcome-based best practices and identifying transactions that most strongly exhibit the characteristics associated with wash trading, our approach reinforces its role as a critical complement to, rather than a substitute for, due process and nuanced legal adjudication.

In sum, this paper advances the literature by offering a rigorous, multi-layered analysis that not only challenges existing methodologies but also provides innovative solutions for accurately detecting wash trading in NFT markets and beyond. By critically evaluating traditional approaches and integrating state-of-the-art machine learning techniques, we deliver a scalable framework that enhances our understanding of digital asset manipulation and informs both academic inquiry and practical regulatory policymaking. As the Web3 ecosystem continues to evolve, future research should extend these methodologies to broader asset classes, ensuring that regulatory measures keep pace with the rapid innovations reshaping financial markets.

\bibliography{main}
\bibliographystyle{agsm}

\appendix
\newpage 
\section{ML Hyperparameters used for training}\label{sec:hp}

\begin{table}[H]
\centering
\begin{tabular}{l|l|l|l|l}
\hline
\textbf{Parameter} & \textbf{Mt. Gox} & \textbf{Blur} & \textbf{LooksRare} & \textbf{OpenSea} \\
\hline
n\_d & 256 & 256 & 256 & 128 \\
n\_a & 256 & 256 & 256 & 128 \\
n\_steps & 1 & 1 & 1 & 1 \\
gamma & 1.4 & 1.4 & 1.4 & 1.4 \\
lambda\_sparse & 1e-4 & 1e-4 & 1e-4 & 1e-4 \\
optimizer & Adam & Adam & Adam & Adam \\
learning\_rate & 2e-2 & 2e-2 & 2e-2 & 2e-2 \\
scheduler & StepLR & StepLR & StepLR & StepLR \\
step\_size & 10 & 10 & 10 & 10 \\
scheduler\_gamma & 0.92 & 0.92 & 0.92 & 0.92 \\
mask\_type & sparsemax & sparsemax & sparsemax & sparsemax \\
momentum & 0.2 & 0.2 & 0.2 & 0.2 \\
batch\_size & $2^{15}$ & $2^{15}$ & $2^{15}$ & $2^{15}$ \\
virtual\_batch\_size & 8000 & 8000 & 8000 & 8000 \\
num\_workers & 24 & 24 & 24 & 24 \\
max\_epochs & 10000 & 10000 & 10000 & 10000 \\
patience & 500 & 500 & 500 & 500 \\
device & cuda & cuda & cuda & cuda \\
\midrule
avg. runtime & overnight & overnight & overnight & overnight \\
\hline
\end{tabular}
\caption{Final hyperparameters used for DL}
\end{table}

\begin{table}[H]
\centering
\begin{tabular}{l|l|l|l|l}
\hline
\textbf{Parameter} & \textbf{Mt. Gox} & \textbf{Blur} & \textbf{LooksRare} & \textbf{OpenSea} \\
\hline
tree\_method & hist & hist & hist & hist \\
device & cuda & cuda & cuda & cuda \\
n\_jobs & -1 & -1 & -1 & -1 \\
n\_estimators & 110 & 100 & 100 & 5000 \\
max\_depth & 15 & 5 & 5 & 15 \\
learning\_rate & 0.09 & 0.1 & 0.1 & 0.1 \\
subsample & 0.9 & 0.8 & 0.8 & 0.8 \\
colsample\_bytree & 0.9 & 0.8 & 0.8 & 0.8 \\
min\_child\_weight & 2 & 1 & 1 & 1 \\
alpha & 0.1 & 0 & 0 & 0 \\
reg\_lambda & 1.5 & 1 & 1 & 1 \\
random\_state & 42 & 42 & 42 & 42 \\
scale\_pos\_weight & 1 & Computed & Computed & Computed \\
early\_stopping\_rounds & 50 & None & None & None \\
eval\_metric & auc & auc & auc & auc \\
\midrule
avg. runtime & minutes & minutes & minute & minutes \\ 
\hline
\end{tabular}
\caption{Final hyperparameters used for XGB}
\end{table}

\begin{table}[H]
\centering
\begin{tabular}{l|l|l|l|l}
\hline
\textbf{Parameter} & \textbf{Mt. Gox} & \textbf{Blur} & \textbf{LooksRare} & \textbf{OpenSea} \\
\hline
n\_estimators & 100 & 100 & 100 & 100 \\
max\_depth & 5 & 5 & 5 & 5 \\
max\_features & sqrt & sqrt & sqrt & sqrt \\
n\_jobs & -1 & -1 & -1 & -1 \\
random\_state & 42 & 42 & 42 & 42 \\
class\_weight & balanced & balanced & balanced & balanced \\
\midrule
avg. runtime & minutes & minutes & minutes & minutes \\ 
\hline
\end{tabular}
\caption{Final hyperparameters used for RF}
\end{table}

\end{document}